\documentclass[twocolumn]{aastex6}
\usepackage{graphicx,epstopdf}
\usepackage{subfigure}
\usepackage{rotating}
\usepackage{longtable}

\usepackage{txfonts}
\begin{document}
\renewcommand{\abstractname}{}    
\title{Statistical Characterization of Hot Jupiter Atmospheres Using Spitzer's Secondary Eclipses}

\author{Emily~Garhart\altaffilmark{1,2,3}, Drake Deming\altaffilmark{1} }
\altaffiltext{1}{Department of Astronomy University of Maryland at College Park, College Park, MD 20742, USA}
\altaffiltext{2}{School of Earth and Space Exploration, Arizona State University, Tempe, AZ 85287, USA}
\altaffiltext{3}{NSF Graduate Research Fellow}

\author{Avi~Mandell\altaffilmark{4} }
\altaffiltext{4}{Planetary Systems Laboratory, Code 693, NASA's Goddard Space Flight Center, Greenbelt, MD 20771, USA}

\author{Heather~A.~Knutson\altaffilmark{5} }
\altaffiltext{5}{Division of Geological and Planetary Sciences, California Institute of Technology, Pasadena, CA 91125, USA}

\author{Nicole~Wallack\altaffilmark{5} }

\author{Adam~Burrows\altaffilmark{6} }
\altaffiltext{6}{Department of Astrophysical Sciences, Princeton University, Princeton, NJ 08544, USA}

\author{Jonathan~J.~Fortney\altaffilmark{7} }
\altaffiltext{7}{Department of Astronomy and Astrophysics, University of California, Santa Cruz, CA 95064, USA}

\author{Callie~Hood\altaffilmark{7} }

\author{Christopher~Seay\altaffilmark{7} }

\author{David~K.~Sing\altaffilmark{8} }

\altaffiltext{8}{Department of Physics and Astronomy, Johns Hopkins University, Baltimore, MD 21218, USA }

\author{Bj\"orn~Benneke\altaffilmark{9} }
\altaffiltext{9}{Departement de Physique, Universite de Montreal, Montreal, H3T 1J4, Canada }

\author{Jonathan~D.~Fraine\altaffilmark{10} }
\altaffiltext{10}{Space Telescope Science Institute, 3700 San Martin Drive, Baltimore, MD 21218, USA}

\author{Tiffany~Kataria\altaffilmark{11} }
\altaffiltext{11}{Jet Propulsion Laboratory, California Institute of Technology, 4800 Oak Grove Drive, Pasadena, CA 91109, USA }

\author{Nikole~Lewis\altaffilmark{10, 12} }
\altaffiltext{12}{Department of Astronomy and Carl Sagan Institute, Cornell University, 122 Sciences Drive, Ithaca, NY 14853, USA }

\author{Nikku~Madhusudhan\altaffilmark{13} }
\altaffiltext{13}{Institute of Astronomy, University of Cambridge, Cambridge, CB3 0HA, UK }

\author{Peter~McCullough\altaffilmark{8, 10}}

\author{Kevin~B.~Stevenson\altaffilmark{10, 14}}
\altaffiltext{14}{JHU Applied Physics Laboratory, 11100 Johns Hopkins Rd, Laurel, MD 20723, USA}

\author{Hannah~Wakeford\altaffilmark{10}}

\shorttitle{Eclipses of Hot Jupiters}
\begin{abstract}

We report 78 secondary eclipse depths for a sample of 36 transiting hot Jupiters observed at 3.6- and 4.5\,$\mu$m using the Spitzer Space Telescope.  Our eclipse results for 27 of these planets are new, and include highly irradiated worlds such as KELT-7b, WASP-87b, WASP-76b, and WASP-64b, and important targets for JWST such as WASP-62b.  We  find that WASP-62b has a slightly eccentric orbit ($e\cos{\omega} = 0.00614\pm{0.00064}$), and we confirm the eccentricity of HAT-P-13b and WASP-14b. The remainder are individually consistent with circular orbits, but we find statistical evidence for eccentricity increasing with orbital period in our range from 1 to 5 days. Our day-side brightness temperatures for the planets yield information on albedo and heat redistribution, following \citet{cowan}.  Planets having maximum day side temperatures exceeding $\sim2200$K are consistent with zero albedo and distribution of stellar irradiance uniformly over the day-side hemisphere. Our most intriguing result is that we detect a systematic difference between the emergent spectra of these hot Jupiters as compared to blackbodies. The ratio of observed brightness temperatures, Tb(4.5)/Tb(3.6), increases with equilibrium temperature by $100\pm24$ parts-per-million per Kelvin, over the entire temperature range in our sample (800K to 2500K). No existing model predicts this trend over such a large range of temperature.  We suggest that this may be due to a structural difference in the atmospheric temperature profile between the real planetary atmospheres as compared to models.
\end{abstract}

\section{Introduction} \label{sec_intro}
The secondary eclipse of a transiting planet provides an opportunity to measure the planet's emitted thermal flux in the infrared spectral region \citep{charbonneau05, deming05}.  When measured over multiple bands, that flux can be used to infer the emergent spectrum of the planet, and numerous investigations have observed and analyzed eclipse photometry for that purpose using the Spitzer Space Telescope (e.g., \citealp{charbonneau08, knutson09}; for a recent review see \citealp{alonso}).  Ideally, the eclipse could be measured spectroscopically with \textit{Spitzer}, but {\it Spitzer's} modest aperture has collected sufficient light to allow eclipse spectroscopy for only two of the brightest hot Jupiter systems \citep{richardson07, grillmair08, todorov14}.  Emergent spectra of several hot Jupiters have been measured near 1.4\,$\mu$m wavelength using the Hubble Space Telescope \citep{kreidberg14, beatty17, cartier17, sheppard17, stevenson17, arcangeli18, kreidberg18, mansfield18, nikolov18}.  The James Webb Space Telescope is projected to obtain emergent spectra for numerous hot Jupiters \citep{greene16, stevenson16, bean18}, enabling a major advance in our understanding of their atmospheric physics and chemistry.  

In this paper, we set the stage for JWST eclipse spectroscopy of hot Jupiters by reporting a statistical analysis of 27 new hot Jupiters observed in eclipse at both 3.6\,$\mu$m and 4.5\,$\mu$m using \textit{Spitzer}.  We are currently engaged in a uniform re-analysis of the secondary eclipses of {\it all} transiting planets observed by \textit{Spitzer}.  A full report on that re-analysis is not yet possible, so we here apply our uniform analysis to hot Jupiters that have not been previously observed or analyzed in secondary eclipse, supplemented by re-analysis of a few planets that either have special and timely interest, such as HAT-P-13b \citep{buhler16, hardy17}, KELT-2Ab \citep{piskorz18}, and WASP-18b \citep{sheppard17, arcangeli18}, or help us to check our eclipse depths in a statistical sense, such as WASP-14b \citep{wong15}.   Given recent interest in the hottest of the hot Jupiters \citep{haynes15, bell17, bell19, evans17, sheppard17, stevenson17, arcangeli18, kreidberg18, mansfield18}, we have tried to be as complete as possible for the hottest planets.  JWST observations of these planets at secondary eclipse will require knowing the orbital phase of their eclipses.   Moreover, slightly non-zero eccentricities for the orbits of hot Jupiters, as revealed by the phase of the secondary eclipse, can be diagnostic of their orbital and physical evolution.  Hence, we also report and discuss the central phase of the eclipses we analyze.  Our work here represents the largest collection of \textit{Spitzer's} secondary eclipse depths ever reported in a single paper.

This paper is organized as follows.  We describe our observations and photometry procedures in Section~\ref{sec_obs}. Section~\ref{sec_extract} describes the analysis of the data, beginning with {\it transits} of three planets to update their orbital periods (Section~\ref{ssec_ephem}). Sections~\ref{ssec_depths} and \ref{ssec_finding} derive eclipse depths and orbital phases by applying pixel-level decorrelation (PLD) to the photometry \citep{deming15}.   Section~\ref{ssec_checks} describes some checks that we have performed to validate our eclipse depths. The eclipse depths of some planets must be corrected for the presence of close companion stars, and those corrections are described in Section~\ref{ssec_dilution}.  Section~\ref{sec_phase} discusses the observed phases of the eclipses, and the implications for orbital dynamics and also for the exoplanetary atmospheres.  Section~\ref{sec_bright} describes how we convert the eclipse depths to brightness temperatures, that are used in the remainder of the analyses.  Sec~\ref{sec_heat} uses those brightness temperatures to study the re-distribution of heat on the planets, and Section~\ref{sec_spectra} compares our measured brightness temperatures to theoretical emergent spectra of the planets. Section~\ref{sec_summary} summarizes our results and conclusions.  An Appendix gives notes on individual planets.

\section{Observations and Photometry} \label{sec_obs}

The bulk of our observations were made under \textit{Spitzer} programs 10102, 12085, and 13044 (PI: Drake Deming) in the 2014-2017 time period. We supplement those observations using archival data for planets observed under other programs. Table~\ref{eclipse_depths} lists the planets we analyze, and the Astronomical Observation Request (AOR) number of each eclipse. Every planet was analyzed using post-cryogenic 3.6- and 4.5\,$\mu$m data from the IRAC instrument.  Most planets were observed in subarray mode, yielding 32x32-pixel images in cubes of 64 frames.  In addition to observations of secondary eclipses, our Cycle-13 program included observations of {\it transits} for many planets.  Analysis of the transits is relevant to transmission spectroscopy of these planets, many of which are being observed by HST/WFC3.  Although this paper focuses on secondary eclipses, we analyze transits of three planets (Sec.~\ref{ssec_ephem}) in order to improve their orbital ephemerides and thereby derive more accurate secondary eclipse phases.
 
To perform photometry, we first remove hot pixels in each frame through a 4$\sigma$ rejection applied to each pixel as a function of time.  We replace bad pixels with the median value of that pixel over time (see \citealp{tamburo18} for a discussion of this median-replacement procedure). We estimate the background by first masking the star with a 5x5 pixel box and tabulating the distribution of pixel intensities outside of this box. The center of a Gaussian fit to this distribution is used as the background value. The code produces photometry by first locating the center of the stellar image on the cleaned 32x32 pixel frame with a 2D Gaussian fit. This initial estimate is refined by two methods: a second 2D Gaussian fit or a center-of-light method. The second Gaussian fit is performed on a smaller (4x4 pixel) box surrounding the initial estimate of the centroid. The center of light position is found with an intensity-weighted average of the X and Y positions nearest the initial estimate. 

We use the \emph{aper} procedure in the IDL's Astronomy User Library to perform the actual aperture photometry, with both fixed-radius and variable-radius apertures methods. Our fixed aperture radii are incremented by 0.1 or 0.2 pixels from 1.6 to 3.5 pixels, producing 11 sets of photometry. The variable radii are computed using the noise-pixel parameter, $\sqrt{\beta}$ from \citet{lewis13}, added to a constant that ranges from 0.0 to 2.0 pixels, depending on the aperture set of the photometry. The combination of two centering methods, and two aperture radii sets, produces a total of four photometric versions of the secondary eclipse for each visit to a given system.  Each version encompasses multiple sets of photometry with different aperture radii.  Each photometric point has an associated time extracted from the headers of the FITS files, as BJD(UTC).  We carry the UTC-based times through the analysis, and subsequently convert the times of the fitted eclipses to TDB, following \citet{eastman10}.

\section{Extraction of Secondary Eclipse Parameters} \label{sec_extract}

\subsection{Ephemeris Updates} \label{ssec_ephem}

\begin{figure}[ht!]
\plotone{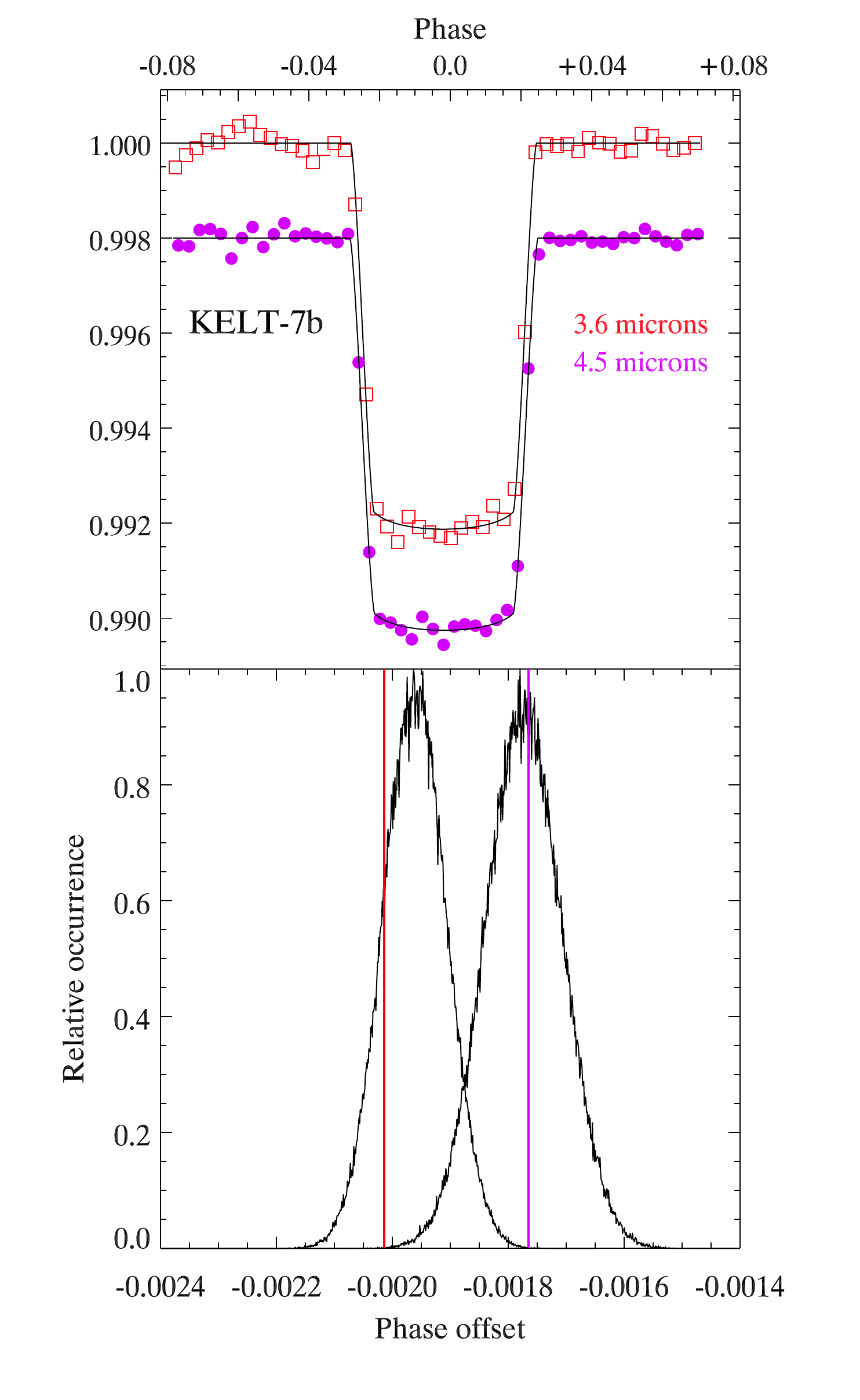}
        \caption{\textit{Spitzer} transits of KELT-7b. The top panel shows the fit to the data, that are here binned so that 50 points span the range of the data, for clarity and consistency of the illustration. (Our fits were carried out using an alternative binning selected by our code). The lower panel shows the likelihood distributions for the central phase; note that they are significantly offset from the predicted phase of zero, but the two \textit{Spitzer} wavelengths are in agreement to within the errors. The vertical lines mark the phase of the best fits selected by our MCMC code, see Section~\ref{ssec_depths}.  For this plot, we use the ephemeris given in the discovery paper by \citet{bieryla15}; see Table~\ref{ephem_table} for transit times and our updated ephemeris.}
\label{KELT7_transit_fig}        
\end{figure}

\begin{figure}[htp]
\plotone{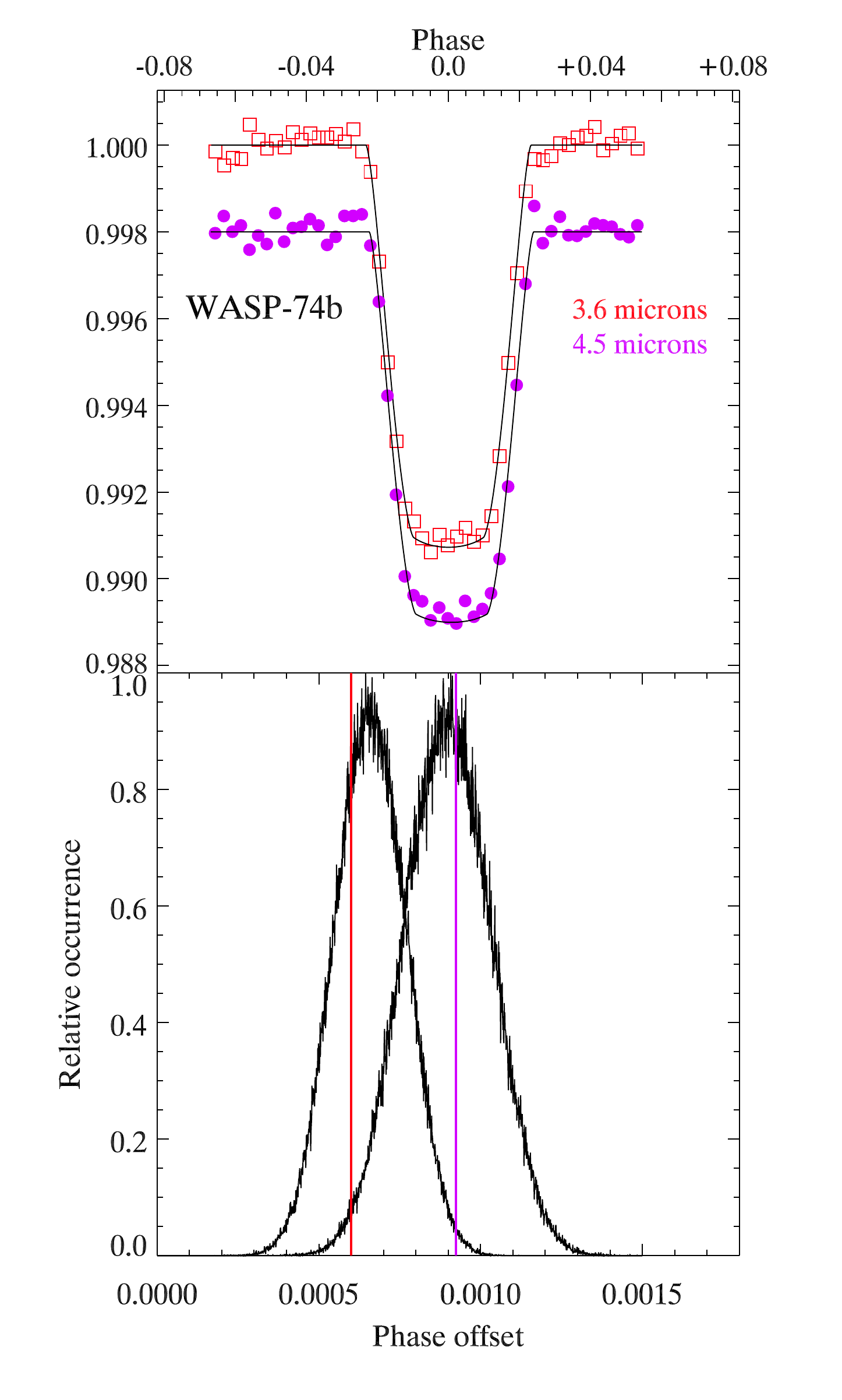}
        \caption{\textit{Spitzer} transits of WASP-74b. The top panel shows the fit to the data, 
which are binned so that 50 points span the range of the data for clarity and consistency of the illustration. 
(Our fits were carried out using an alternative binning selected by our code).  The lower panel shows the likelihood distributions for the central phase; note that they are significantly offset from the predicted phase of zero, but the two \textit{Spitzer} wavelengths are in agreement to within the errors. The vertical lines mark the phase of the best fits selected by our MCMC code, see Section~\ref{ssec_depths}. 
For this plot, we use the ephemeris given in the discovery paper by \citet{hellier15}; see Table~\ref{ephem_table} for transit times and our updated ephemeris.}
\label{W74_transit_fig}        
\end{figure}

\begin{figure}[htp]
\plotone{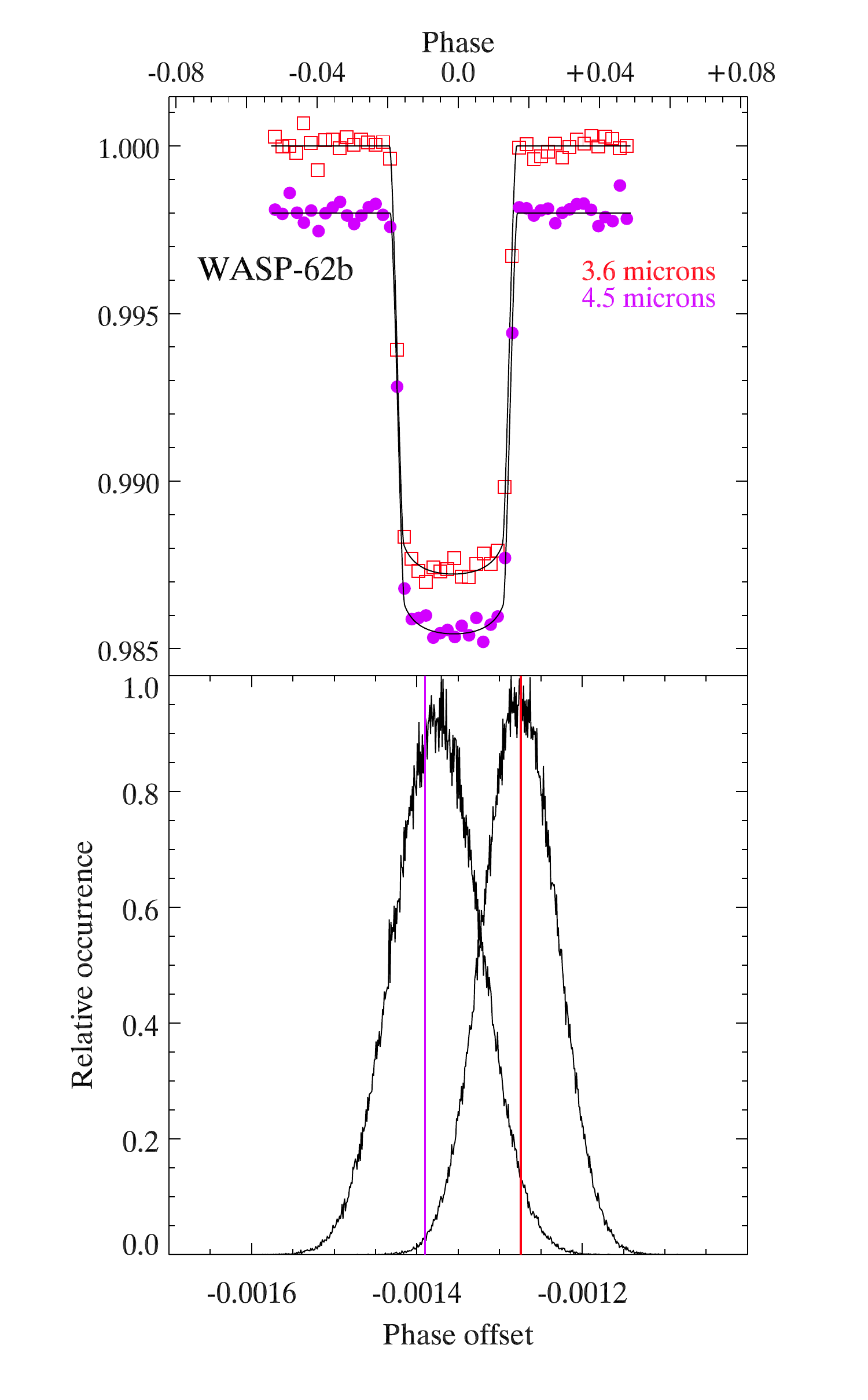}
        \caption{\textit{Spitzer} transits of WASP-62b.  The top panel shows the fit to the data, that are here binned so that 50 points span the range of the data, for clarity and consistency of the illustration. (Our fits were carried out using an alternative binning selected by our code). The lower panel shows the likelihood distributions for the central phase; note that they are significantly offset from the predicted phase of zero, but the two \textit{Spitzer} wavelengths are in agreement to within the errors. The vertical lines mark the phase of the best fits selected by our MCMC code, see Section~\ref{ssec_depths}.  For this plot, we use the ephemeris given in the discovery paper by \citet{hellier12}; see Table~\ref{ephem_table} for transit times and our updated ephemeris. }.
\label{W62_transit_fig}        
\end{figure}

The time scale for tidal circularization of a hot Jupiter's orbit is typically much less than the age of the system \citep{jackson08}.  Observations commonly find hot Jupiter secondary eclipses to be centered very close to phase 0.5 (e.g., \citealp{garhart18}), consistent with a circular orbit.  When we find a displacement of the eclipse from phase 0.5, we first check the impact of potential ephemeris error on the observed phase of the eclipse.  We found three planets whose ephemerides we were able to update: KELT-7b, WASP-62b, and WASP-74b.
We fit \textit{Spitzer} transits for each planet at both 3.6- and 4.5\,$\mu$m using the same procedure as for our eclipse fits (See Section~\ref{ssec_depths} below), except that we include quadratic limb darkening based on coefficients in each band from \citet{claret13}. We freeze the orbital parameters and limb darkening coefficients during the fit, and we vary the ratio of radii (planet-to-star) and the  central phase of the transit. Low infrared limb darkening produces a sharp ingress/egress for the \textit{Spitzer} transits, and facilitates a precise measurement of the transit time. For KELT-7b and WASP-74b, we find that the \textit{Spitzer} transits are displaced from their predicted phases by amounts that are consistent between the two \textit{Spitzer} bandpasses, and commensurate with the offsets we encountered for the eclipses.  The observed transits and fits are illustrated in Figures~\ref{KELT7_transit_fig}, ~\ref{W74_transit_fig} and~\ref{W62_transit_fig}.  The transit times are given in Table~\ref{ephem_table}, and the transit depths are given in Table~\ref{tdepth_table}.

We update the orbital periods of KELT-7b and WASP-74b using the \textit{Spitzer} transit times.  For each planet, we use the transit epoch ($T_0$) from \citet{bieryla15} and \citet{hellier15}, and we calculate a new period using three points: the epoch listed in the discovery paper, and the transit times from our new \textit{Spitzer} transits (one at each wavelength).  We calculate the period via error-weighted linear least-squares (\emph{linfit} routine in IDL), and the error on the slope (i.e., the period) follows from the precision of the original $T_0$ value and the precision of the \textit{Spitzer} transit times. The precision of the updated period for KELT-7b is improved by a factor of 8 compared to \citet{bieryla15}, and for WASP-74b by a factor of 2 compared to \citet{hellier15}.  The \textit{Spitzer} transit times and updated periods are given in Table~\ref{ephem_table}, and those values are used to calculate the secondary eclipse phases reported in this paper (Section~\ref{sec_phase}). 

For WASP-62b, the transits are similarly displaced slightly from the predicted time, as shown on Figure~\ref{W62_transit_fig}.  Again, there is excellent agreement between the transits measured independently in both \textit{Spitzer} bands.  We have updated the ephemeris based on the \textit{Spitzer} transits, and the updated results are included in Table~\ref{ephem_table}.  However, even with our updated ephemeris, the eclipses of WASP-62b remain displaced from phase 0.5 due to an eccentric orbit, as discussed in Section~\ref{sec_phase}. 

\subsection{Analyzing the Photometry} \label{ssec_depths}

Two major instrumental systematic effects are known to contaminate \textit{Spitzer} observations and introduce fluctuations in the photometry that can often be an order of magnitude larger than the eclipse being sought. First, there is a characteristic ramping feature that varies with time \citep{knutson12}. This ramp-like increase in flux is often most rapid at the beginning of each observation, so by default we omit the first 30 minutes of data from every eclipse to eliminate the potentially steepest portion of the ramp.  We model the ramp in the remaining data using either a linear, quadratic, or exponential function of time.  We decide between a linear and quadratic ramp model using a Bayesian Information Criterion (BIC) applied to the fitted eclipse.  In the (infrequent) cases where the fit is inadequate near the beginning of the time series (judged by structure in the residuals), we either omit 45 or 60 minutes of data instead of the default 30 minutes, or we use an exponential ramp, depending on the characteristics of those specific data. 

The second source of noise for \textit{Spitzer} is the intra-pixel sensitivity variations across the detector. We correct for this effect using Pixel Level Decorrelation (PLD, \citealp{deming15}), and including the temporal ramp as integral to the PLD fitting process.  In a \textit{Spitzer} data challenge, \citet{ingalls16} found PLD to have the smallest bias in eclipse measurements as compared to other current decorrelation methods. PLD has been extensively used for \textit{Spitzer} analyses \citep{dittmann17, kilpatrick17, buhler16, fischer16, wong16, tamburo18}, and higher-order PLD is the foundation of the EVEREST code for analysis of K2 photometry \citep{luger16}. The PLD formalism was described by \citet{deming15}, and we do not repeat the equations here.  But we summarize that the photometry is modeled as proportional to a linear sum of normalized relative pixel intensities times coefficients determined by the fit, and including the temporal ramp and the eclipse shape.  Moreover, our PLD fit uses binned data, because binning averages out small temporal scale fluctuations in the basis pixels, and reduces or eliminates red noise much more efficiently than with unbinned data.  Normalizing the pixels is used to remove all astrophysical information from the independent variables in the fitting process.  We calculate the shape of the eclipse with an adapted version of the procedure described by \citet{mandel}.   As described by \citet{garhart18}, our version of PLD uses 12 basis pixels, versus the original 9 pixels used by \citet{deming15}. These 12 pixels are the closest to the median stellar center found in the photometry and generally form a 4x4 pixel box without corners.  The eclipse depth is not sensitive to the number of basis pixels {\it per se}, but the stars in our sample are sufficiently bright on average that significant flux can be detected in more than the 9 pixels originally used by \citet{deming15}, and we want to use all significant pixel-level information. Note that \citet{tamburo18} used 25 basis pixels for the very bright star 55~Cnc. 

\subsection{Finding the Eclipse Depth} \label{ssec_finding}

We here describe the specific procedure that we use when fitting the data to determine the best secondary eclipse depths.  Our method is formally Bayesian, but we use uniform priors (see below), so in practice it reduces to a maximum-likelihood calculation.  We use the best available orbital parameters for each planet, but we freeze them during the fitting process, varying only the eclipse depth and central phase. That has ample precedent based on many previous secondary eclipse investigations (e.g., \citealp{orourke14, evans15, mansfield18}).  We use a uniform prior on the eclipse phase that covers the range of the {\it Spitzer} observations for each planet.

The duration as well as the phase of a secondary eclipse is affected by non-zero orbital eccentricities \citep{charbonneau05}.  Anticipating our eclipse phase results (Sec.~\ref{sec_phase}) that constrain eccentricities, and using Eq.~(5) of \citet{charbonneau05}, we calculate that the difference between transit and eclipse duration is not detectable given our precision on the observed eclipses.  For planets with known non-zero orbital eccentricities, we nevertheless account for the effect of the eccentricity on the modeled eclipse durations.

An alternative to freezing the orbital parameters would be to use a Gaussian prior for each orbital parameter. However, when those priors are independent of each other, our MCMC (see below) could step to regions of parameter space that would not be acceptable when fitting the transit data, which have much higher S/N than eclipses.  For example, the orbital inclination and $a/R_s$ are correlated when fitting transits because they both affect the transit duration, and can trade-off against each other.  Using uncorrelated Gaussian priors when fitting eclipses allows combinations of inclination and $a/R_s$ that are not constrained {\it by the actual transit data}. So in those cases there is the danger of using orbital parameters that the transit data would reject. It isn't practical to fit all of the transit data simultaneously with the large number of \textit{Spitzer} eclipses that we analyze here.  Therefore we continue to fit the eclipses by freezing the best orbital parameters, and varying only the eclipse depth and central phase.  

We find the best-fitting eclipse depth and central phase by maximizing the Bayesian posterior probability of each hypothetical depth value, given as:
\begin{equation}
    \mathcal{P}_{posterior} \propto \mathcal{L} \times \mathcal{P}_{prior},
\end{equation}
where the $\mathcal{P}$ values are the prior and posterior probabilities, and $\mathcal{L}$ is the likelihood, given as:
\begin{equation}
    L = \prod_{i=1}^{n}(1/\sqrt(2{\pi}{\sigma}^2))\exp(-(m_i-d_i)/2\sigma^2),
\end{equation}
where $d_i$ are the data ($n$ photometry points), $m_i$ are the modeled photometry points (sum of the eclipse model and the PLD model of the intra-pixel detector sensitivity, see \citealp{deming15}).  ${\sigma}$ is the uncertainty assigned to the data points (the same for all points in a given eclipse).  Note that Eq.(1) is essentially Bayes' Theorem, without the denominator (Bayesian evidence, a constant) on the right hand side.  Because we freeze the orbital parameters except for eclipse phase, we have:
\begin{equation}
  \mathcal{P}_{prior} = 1/ (\phi_{max}-\phi_{min}),  
\end{equation}
where $\phi_{max}$ and $\phi_{min}$ are the maximum and minimum orbital phases present in the data. For a given eclipse, $\mathcal{P}_{prior}$ is a constant, and our code maximizes $\mathcal{P}_{posterior}$ by maximizing $\ln{\mathcal{L}}$:
\begin{equation}
    \ln{\mathcal{L}} = \sum_{i=1}^{n}{-(m_i-d_i)^2/2\sigma^2} + constants,
\end{equation}
and the right hand side is just the negative of a conventional $\chi^2$, plus constants. So for the eclipses we analyze, maximizing $\mathcal{P}_{posterior}$ reduces to minimizing $\chi^2$. 

Our fitting code uses an initial linear regression to locate the eclipse and estimate the best central phase and pixel coefficients by minimizing the $\chi^{2}$ of a fit to the unbinned data.  Then, we freeze the phase of the eclipse, and re-fit for the \textit{Spitzer} systematics and the eclipse depth using binned data with combinations of aperture radius and bin size, again using linear regression. For each fit to binned data, the code uses the best pixel coefficients and best eclipse depth from the regression to calculate a fit to the unbinned data, and subtracts that to form residuals.  The code then calculates the variance (${\sigma}^2$) of the residuals as a function of bin size (this is called the Allan deviation relation, \citealp{allan66}). We adopt the combination of bin size, aperture type and size, and centering method, that minimizes the scatter in the Allan deviation relation (see \citealp{garhart18}). We allow the code to select negative eclipse depths, to eliminate Lucy-Sweeney bias for weak eclipses \citep{lucy}.

Once the best aperture radius, bin size, and best-fit parameters have been found, they are used to seed a $1 x 10^6$ step Markov Chain Monte Carlo (MCMC) procedure \citep{ford} in order to estimate the errors on both the central phase and eclipse depth.  We separate the MCMC into three distinct stages: an initial burn-in period of approximately $1 x 10^4$ steps on the unbinned data to find the best step sizes for each parameter. After the burn-in, we re-scale the photometric errors so that the reduced $\chi ^{2}$ is $\approx$ 1 for the rest of the analysis. 

Approximately $8 x 10^5$ steps are used to fit the binned data and  adequately sample the entire parameter space as well as to significantly reduce computation time. Finally, the last $1 x 10^4$ steps also calculate the fit to the unbinned data, and re-compute the Allan deviation relation at each step, so as to possibly find a slightly better solution. The MCMC varies the eclipse phase simultaneously with other parameters in this process (whereas the linear regressions held the phase constant after an initial estimate).  Thereby, the MCMC is sometimes able to find a slightly better central phase and eclipse depth value than the linear regressions.  We post-process the MCMC chains to calculate the errors on eclipse depth and central phase by fitting Gaussians to the likelihood distributions from the MCMC, and those are virtually always excellent fits.

As mentioned above, there are four sets of photometry for each wavelength.  We fit the four versions separately and select the best combination of centering method (Gaussian or center-of-light) and aperture type (fixed or variable radii) by considering the ratio of scatter relative to the photon noise, on both the binned and unbinned time scale.  The ratio can vary with bin size, and there is a trade-off between minimizing red noise as opposed to noise on the unbinned time scale.  We have not found a rigid formula to implement this trade-off, so subjective judgment is sometimes needed depending on the characteristics of specific eclipses.  However, we check to ensure that the eclipse depth is not sensitive (within the errors) to the choice, and we also inspect each fitted eclipse visually to check for potential anomalies in the fit.  

In all cases we re-run the code with a different MCMC random seed, to verify convergence to closely similar likelihood distributions of eclipse depth and central phase. Most of our results are based on two MCMC chains per eclipse, each sampling depth and central phase.  In some cases we run four MCMC chains in order to further check convergence.  We compute the Gelman-Rubin statistic, $R$ \citep{gelman}, for all depth and central phase values based on either two or four independent chains per eclipse. \citet{gelman04} suggested that $R$ values less than 1.1 indicate adequate convergence, but exoplanet secondary eclipse work usually achieves values much closer to unity (e.g., \citealp{cubillos14}). Our median $R$ value for the 78 eclipse depths is 1.0023, the average value is 1.0052, and the maximum value is 1.0245.  For central phase, the median, average, and maximum $R$ values are 1.0010, 1.0028, and 1.0276.  

\subsection{Properties and Checks on the Eclipse Solutions}  \label{ssec_checks}

We here describe the properties of our PLD eclipse solutions, and we make a number of checks to ensure the validity of the eclipse depths.  Recall that our PLD fitting process operates on binned data, and chooses a 'broad bandwidth' solution by minimizing the scatter in the Allan deviation relation (see \citealp{garhart18} and Sec.~3.3 of \citealp{deming15}).  We thereby expect that the solutions should be good fits to the data on all time scales, no matter how we bin the data.  For clarity of presentation, we bin the data to between 20 and 40 points spanning each data set, and we show all of the eclipses at 3.6\,$\mu$m in Figure~\ref{all36}, and all of the 4.5\,$\mu$m eclipses in Figure~\ref{all45}.  The eclipse of every planet is nominally detected at 4.5\,$\mu$m (albeit some with low signal-to-noise), and all except for WASP-75b and WASP-49b are detected at 3.6\,$\mu$m (the fitted 3.6\,$\mu$m eclipse has a negative depth for WASP-75b and -49b, indicating that the eclipse amplitudes are beneath the noise). 

\begin{figure}[htp]
\epsscale{1.15}
\plotone{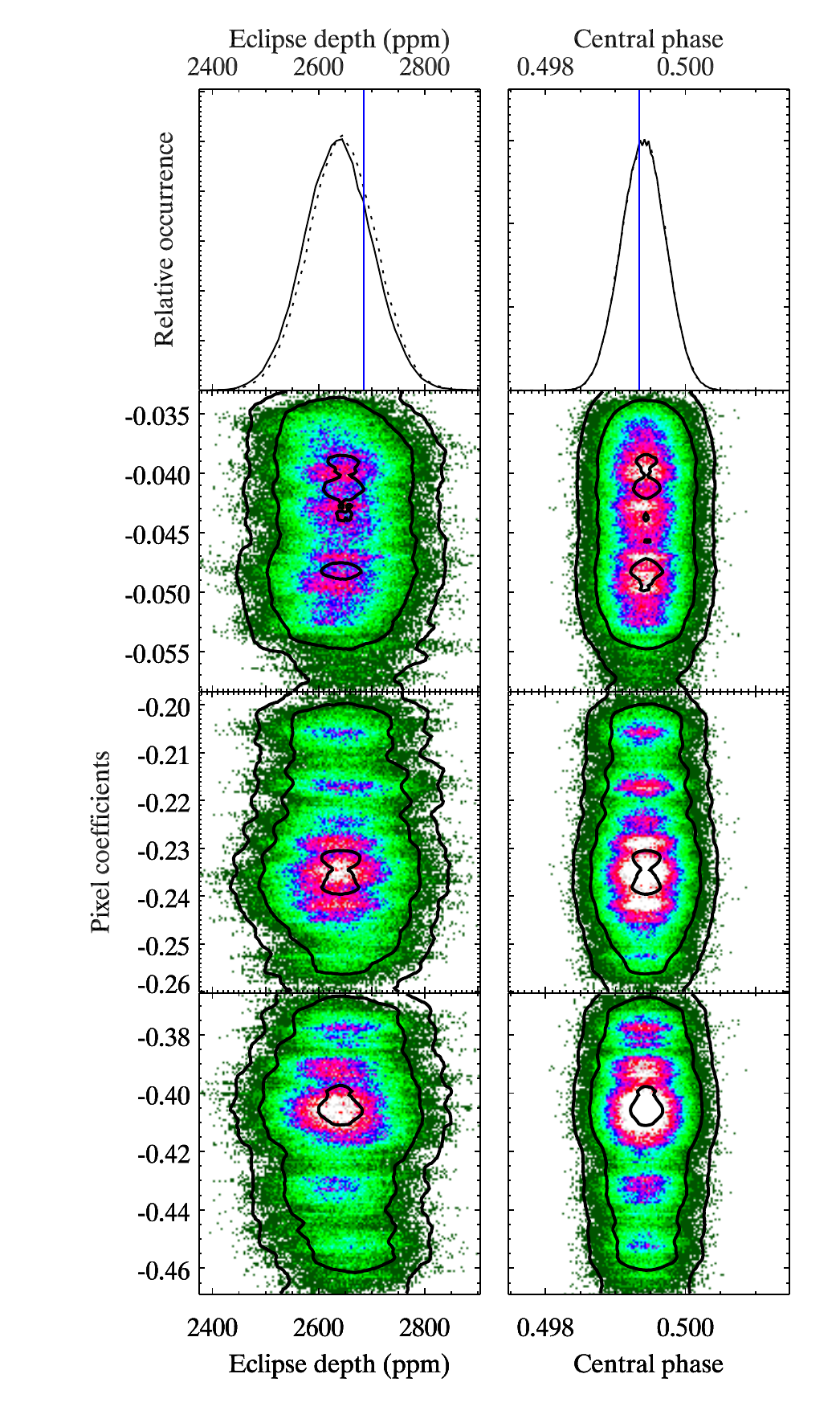}
\caption{Top panel illustrates the distributions for 3.6\,$\mu$m eclipse depth and central phase for WASP-76b. The dashed line (nearly coincident with the solid line) shows nearly identical distributions from duplicate Markov chains with different starting seeds.  The vertical lines are the best-fit values chosen by our code, based on minimizing the scatter in the Allan deviation relation. The three lower panels are the distributions for the three brightest pixels in the PLD solutions, versus the distribution of eclipse depth and central phase. The contours are point densities of 0.01, 0.1, and 0.9 of the maximum density. In all cases, the depth and phase are uncorrelated with the pixel coefficients.}
\label{pix_coeffs_76}
\end{figure}

\begin{figure}[htp]
\epsscale{1.15}
\plotone{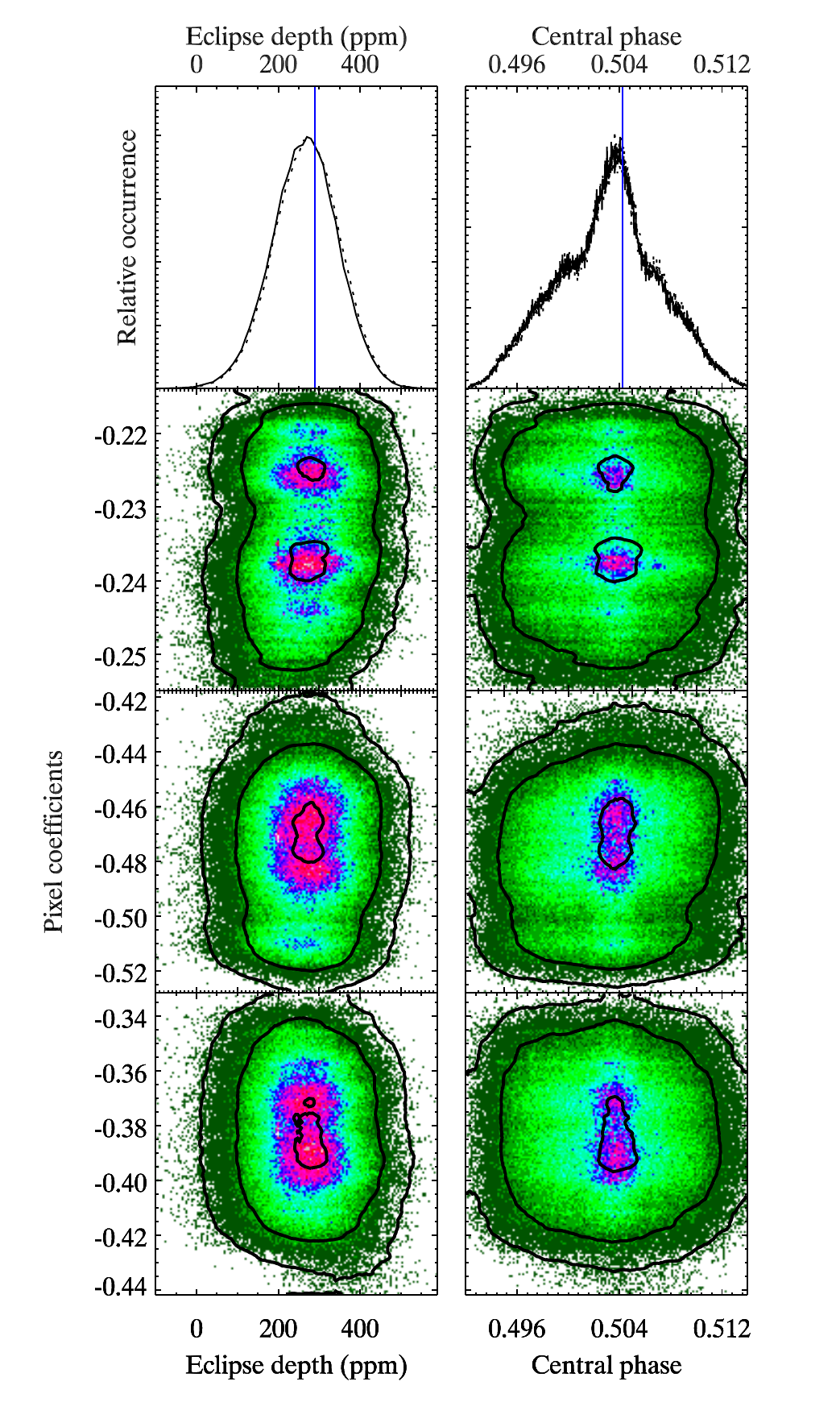}
\caption{Top panel illustrates the distributions for 4.5\,$\mu$m eclipse depth and central phase for WASP-131b. The dashed line (nearly coincident with the solid line) shows nearly identical distributions from duplicate Markov chains with different starting seeds.  The vertical lines are the best-fit values chosen by our code, based on minimizing the scatter in the Allan deviation relation. The three lower panel are the distributions for the three brightest pixels in the PLD solutions, versus the distribution of eclipse depth and central phase. The contours are point densities of 0.01, 0.1, and 0.9 of the maximum density. In all cases, the depth and phase are uncorrelated with the pixel coefficients.}
\label{pix_coeffs_131}
\end{figure}

In addition to the eclipse fits shown in Figures~\ref{all36} and \ref{all45}, we here explore additional properties of the solutions. The arrangement of pixels relative to the position of the stellar image means that the pixel coefficients in the PLD fitting process can correlate and anti-correlate with each other as the stellar image moves.  Given that we expect pixel-to-pixel correlations, a traditional corner plot using the full array of pixel covariances is not particularly useful.  However, the eclipse depth should not strongly correlate with any pixel coefficient, since we expect that the pixels will trade-off appropriately in the presence of a stable eclipse depth as the MCMC evolves.  Accordingly we illustrate the weakness of correlation between the eclipse depth and pixel coefficients, for two representative eclipses, choosing a strong eclipse (WASP-76b) and a weak eclipse (WASP-131b).  Figures~\ref{pix_coeffs_76} and \ref{pix_coeffs_131} show the likelihood distributions for both eclipse depth and central phase, versus the distributions for the three brightest pixel coefficients.  In all cases, no strong correlation is present.  Although we illustrate the three brightest pixels, we calculate Pearson correlation coefficients for all 12 pixels versus the depth of each eclipse for one MCMC chain per eclipse, producing $12 x 78 = 936$ values. The Pearson values measure the significance of possible correlations, and can be positive or negative, so we work with absolute values.  The median Pearson values for the pixel coefficients at 3.6- and 4.5\,$\mu$m are 0.0831 and 0.0829, respectively.  85\% and 83\% of the values are less than 0.2 at 3.6- and 4.5\,$\mu$m, respectively.  Although the Pearson values are small (perfect correlation would produce unity), they can indicate statistically significant correlations in some cases because each Pearson value is based on $8 \times 10^5$ samples in an MCMC chain.  However, the correlations are weak in the sense that their slope is not sufficient to perturb the eclipse depths significantly, especially since the effects trade-off between pixels. 

We track the total effect of the pixel coefficients on the eclipse depth during the evolution of each MCMC chain. We calculate the standard deviation of the effect on eclipse depth, averaged over a time scale of 5000 steps.  The median value of those standard deviations, tabulated over all eclipses, is 2.0\% of the eclipse depths at 3.6\,$\mu$m and 1.5\% at 4.5\,$\mu$m.  We conclude that, although degeneracies between the ramp coefficients and the eclipse depth can contribute significantly to the error on the eclipse depths (see below), the total effect of the pixel coefficients is not significantly degenerate with eclipse depth. 

\begin{figure}[htp]
\epsscale{1.15}
\plotone{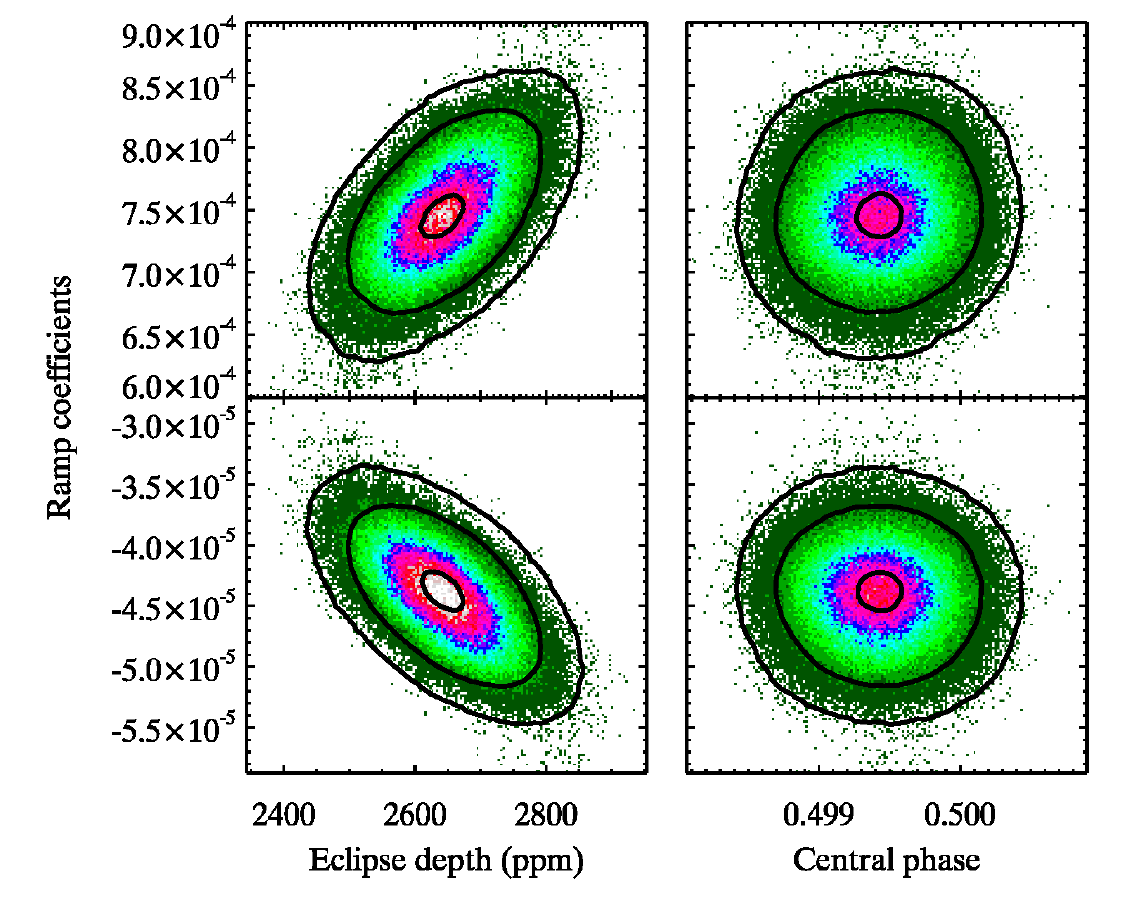}
\caption{Distributions of the temporal ramp coefficients for WASP-76b at 3.6\,$\mu$m, versus the distributions of eclipse depth (left column) and central phase (right column). The top row is the coefficient of $t$ ($t$=time), and the bottom row is the coefficient of $t^2$. The contours are point densities of 0.01, 0.1, and 0.9 of the maximum density.}
\label{ramp_coeffs_76}
\end{figure}

\begin{figure}[htp]
\epsscale{1.15}
\plotone{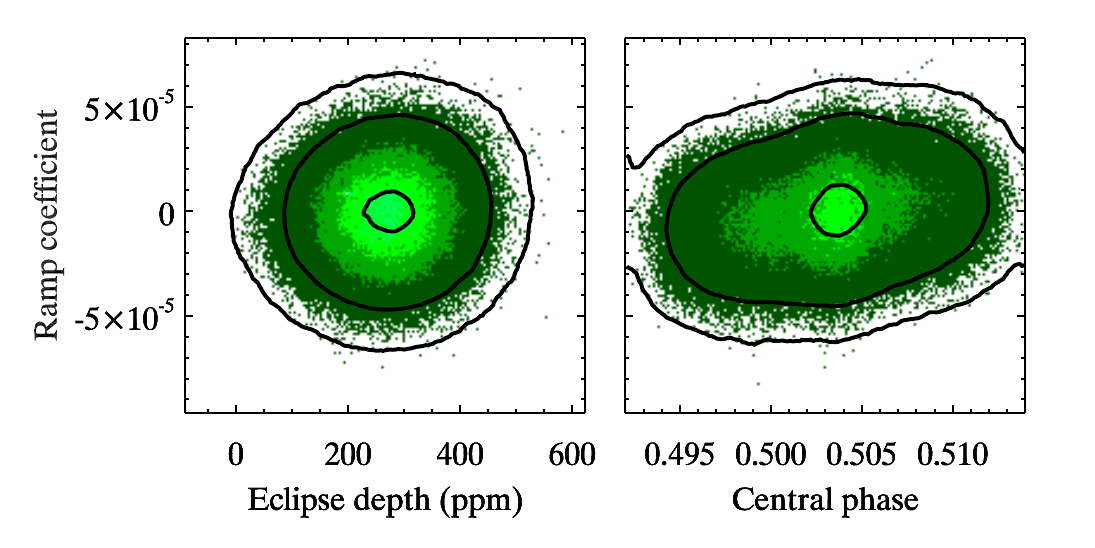}
\caption{Distributions of the temporal ramp coefficient of $t$ (=time) for WASP-131b, versus the distributions of eclipse depth (left) and central phase (right). (Our WASP-131b eclipse used only a linear, not a quadratic ramp).}
\label{ramp_coeffs_131}
\end{figure}

Although the derived eclipse depths and phases do not strongly correlate with the PLD pixel coefficients, they do (and should) correlate with the parameters of the temporal ramp, both for the linear and quadratic case.  That occurs because the presence of a ramp perturbs the out-of-eclipse reference flux, and it also shifts the centroid of the eclipse.  Indeed, the entire point of including the ramp in the solution is to account for such correlations.  Figures~\ref{ramp_coeffs_76} and \ref{ramp_coeffs_131} show those correlations for WASP-76b and -131b, respectively.  The correlations are included in our quoted errors for eclipse depth and central phase (not only for these planets we illustrate but also for all planets we analyze). 

\begin{figure}[htp]
\plotone{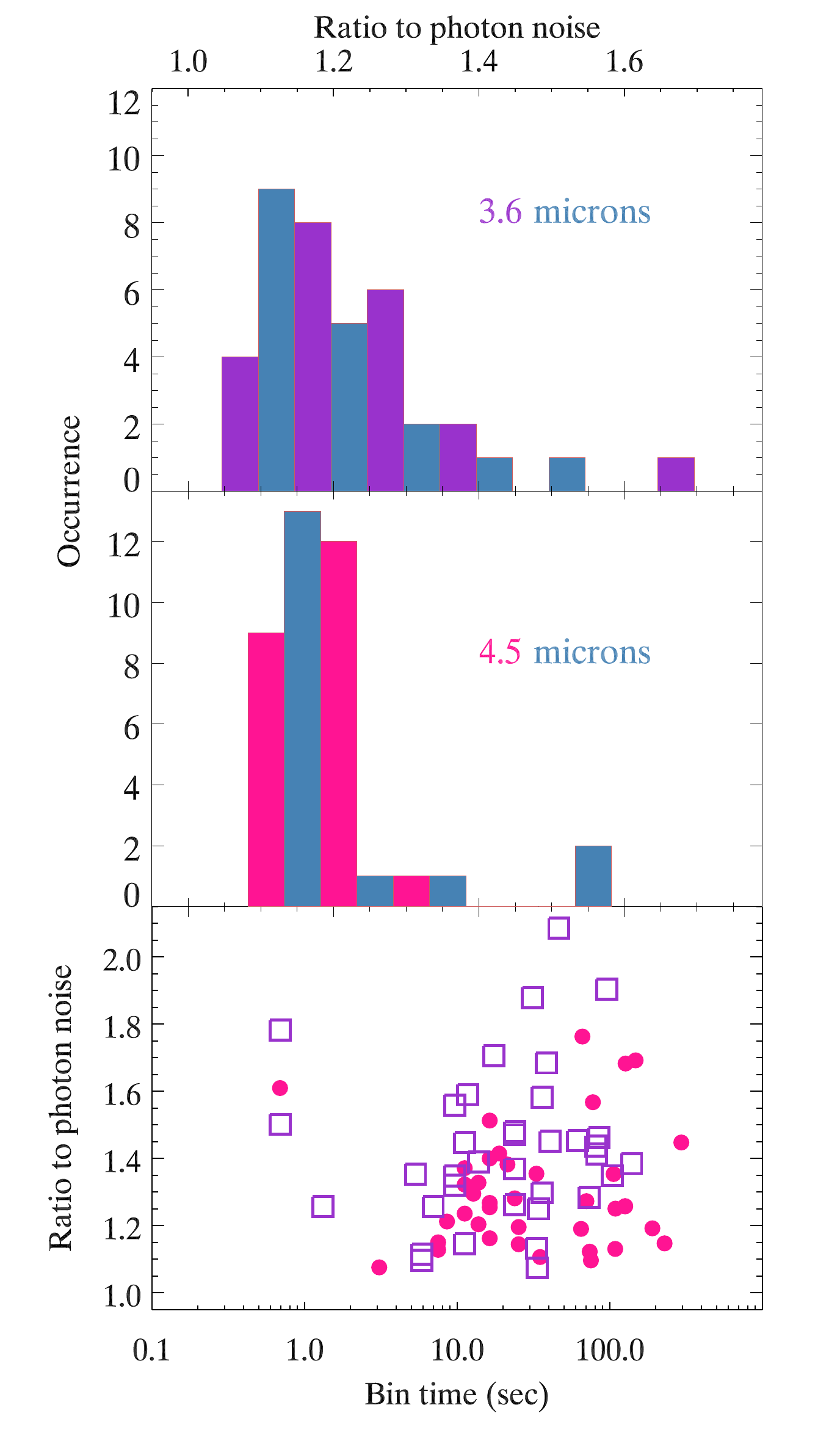}
\caption{The upper two panels (read upper axis scale) show histograms of the ratio of the unbinned scatter in our residuals to the photon noise for all eclipse depth solutions (each planet contributes one point to each histogram).  Alternate colors for adjacent histogram bins are used solely for visual clarity.  The bottom panel (read lower axis scale) shows the ratio of the scatter to the photon noise on the binned time scale used for each eclipse solution, versus the bin time for that solutions. Point colors identify the wavelength, as per the two upper panels. 
 }
\label{ratio_plot} 
\end{figure}

An important check on the properties of our solutions for eclipse depth is to examine the amplitude of the residuals (data minus fit) as a function of bin size.  Recall that our code fits to binned data, because we find that it helps to reduce red noise.  We apply the coefficients from that best fit to the {\it unbinned} data, and subtract that fit.  We re-bin the residuals with a variety of bin sizes, and calculate the scatter (standard deviation, $\sigma$) of each set of binned residuals for both the binned and unbinned data.  Figure~\ref{ratio_plot} shows histograms of this ratio for the unbinned data at both 3.6- and 4.5\,$\mu$m.  The scatter is always greater than the photon noise; at 3.6\,$\mu$m the median ratio is 1.19, and at 4.5\,$\mu$m the median is 1.17.  The distribution at 4.5\,$\mu$m is more strongly concentrated at ratios near unity. At each wavelength, only two eclipses have ratios exceeding 1.5.  The bottom panel of Figure~\ref{ratio_plot} shows the ratio of the scatter to the photon noise on the binned time scales that were actually used for each eclipse solution.  The median values of that ratio are 1.44 and 1.26 at 3.6- and 4.5\,$\mu$m, respectively, but 13 eclipses scatter to ratios above 1.5 at 3.6\,$\mu$m, versus 6 at 4.5\,$\mu$m.  We conclude that the eclipse solutions are giving good performance over a wide range of time scales.   Note also the ratio of scatter to the photon noise does not correlate with the bin time on the bottom panel of Figure~\ref{ratio_plot}, indicating that the scatter is decreasing versus bin size with approximately the same functional behavior for all eclipses.

\begin{figure}[htp]
\plotone{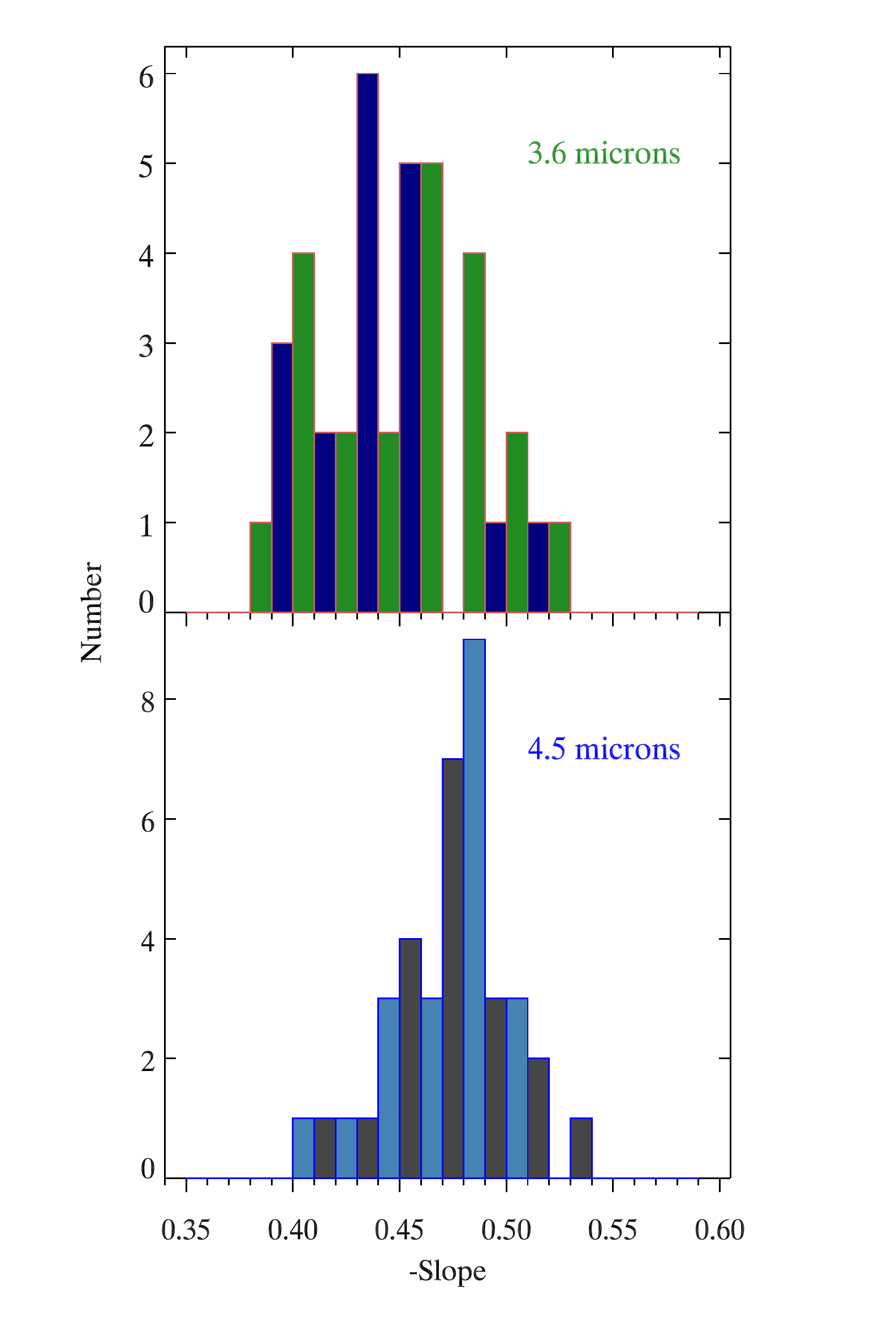}
\caption{Histograms of the Allan deviation slope for our collection of eclipse depth solutions. Alternate colors for adjacent histogram bins are used solely for visual clarity. The top panel shows the distribution at 3.6\,$\mu$m, and the bottom panel at 4.5\,$\mu$m.}
\label{slope_fig} 
\end{figure}

Another way to view the noise performance of the eclipse solutions is from the slope of the Allan deviation relation, i.e. the standard deviation of the binned residuals as a function of bin time.  Histograms of the Allan deviation slope are shown for both wavelengths in Figure~\ref{slope_fig}. For photon-limited performance, the standard deviation ($\sigma$) should decrease as the square root of the bin size with a slope of -0.5 in log space.  If, for example, we were to over-fit the data, then we might find the slope to be consistently less than -0.5, which is not physically possible for a valid fitting process (because we cannot overcome the photon noise).  The distributions of Allan deviation slope over all of our eclipse depth solutions are therefore useful diagnostics of our fitting procedure.  Figure~\ref{slope_fig} shows histograms of the slopes for the 3.6- and 4.5\,$\mu$m eclipses.  The median value for the 3.6\,$\mu$m slopes is -0.45 and for 4.5\,$\mu$m it is -0.48.  Both distributions decrease strongly at -0.5, albeit with some values approaching -0.54.  Our 3.6\,$\mu$m solutions have 4 slope values less than -0.5, but all of them greater than -0.53.  At 4.5\,$\mu$m, 6 slopes are less than -0.5, with the smallest value being -0.537.  The slope has its own intrinsic uncertainty, averaging to 0.013 at both wavelengths. No individual slope is below -0.5 by 3 or more times its individual standard deviation. We conclude that the values falling below -0.5 are due to random fluctuations, and that our eclipse depth solutions approach closely to the photon noise limit, but we are not over-fitting.  

\begin{figure}[htp]
\plotone{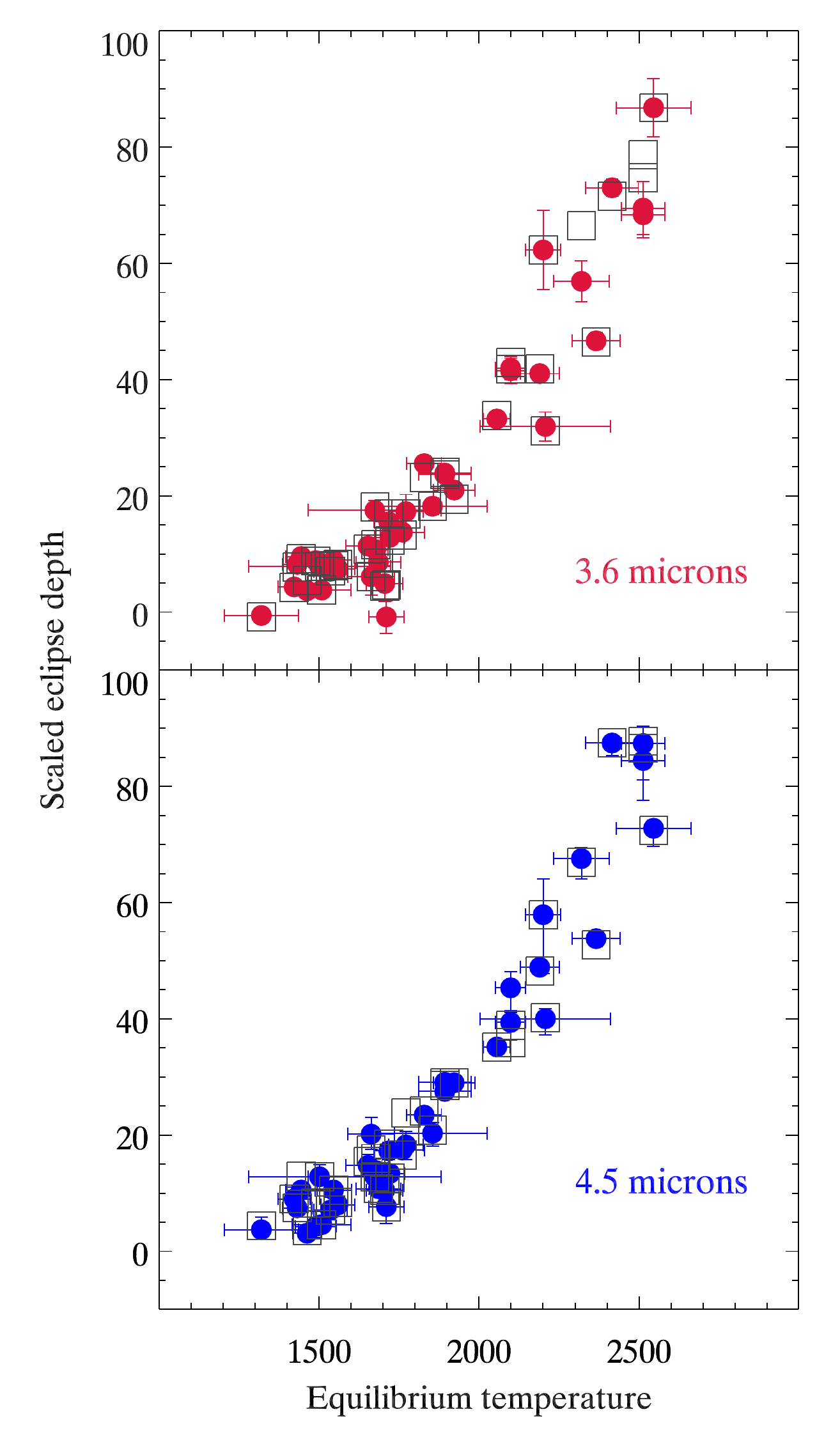}
\caption{Scaled eclipse depths (see text) versus the equilibrium temperature of each planet. The purpose of this comparison is to check the consistency and relative scatter between our adopted eclipse depths (solid points with error bars), and the eclipse depths derived always using Gaussian centroiding and constant-radius photometric apertures (GC depths).  The GC depths are plotted as open squares without error bars.  Both sets of eclipse depths are in excellent agreement, and exhibit similar levels of scatter.}
\label{depths_fig}
\end{figure}

As described above, we examine four different versions of the photometry at each wavelength, independently choosing the best overall fit from among them for each planet and each wavelength.  Thus we might adopt Gaussian centroiding with variable-radius photometry apertures at 3.6\,$\mu$m for a given planet, and center-of-light centroiding with constant-radius apertures for the same planet at 4.5\,$\mu$m.  Our rationale is that each data set is different, and has unique characteristics that require flexibility in the fitting process.  Nevertheless, a strength of our work is that we analyze eclipses for 27 new planets using a uniform methodology, to facilitate accurate statistical conclusions.  In light of that goal, it may seem odd that we utilize one of four different sets of photometry for each planet at each wavelength.  Does this variation destroy the uniformity of our analysis, and introduce additional noise or systematic effects?  To investigate that possibility, we compare our adopted eclipse depths with the eclipse depths that are derived always using Gaussian centroiding and constant-radius apertures (hereafter, Gaussian-constant = GC).  One way to evaluate uniformity is to compare each set of eclipse depths with some physical variable that is independent of our data analysis, but should correlate with eclipse depth. Whatever the shape of that functional relation, the best set of eclipse depths should exhibit less scatter.  We use the equilibrium temperature of each planet as the independent variable, calculated assuming zero albedo, a circular orbit, and uniform distribution of heat.  We remove the effect of different stellar and planetary radii, and the stellar temperature, by dividing each measured eclipse depth (not including the dilution correction described in Sec.~\ref{ssec_dilution}) by the ratio of planetary to stellar disk areas. We also multiply by the stellar intensity, using a blackbody at the stellar effective temperature (a good approximation at these wavelengths). We multiply the result by 100 to put the numbers on a convenient scale.  These scaled eclipse depths are shown at 3.6- and 4.5\,$\mu$m in Figure~\ref{depths_fig}. As expected, both sets of eclipse depths correlate with equilibrium temperature, albeit not a purely linear relation (the exact shape of the relation is unimportant for our immediate purpose).  

Interestingly, the GC eclipse depths yield virtually the same correlation on Figure~\ref{depths_fig}, with the same scatter, as do our adopted eclipse depths.  This shows that we are not introducing a source of significant non-uniformity when choosing from among four different sets of photometry, but neither are we significantly improving the results.  To investigate further, we calculated the linear regression relation between the GC depths and our adopted depths. A maximum likelihood regression (see below) with the adopted depths as Y and GC depths as X yields a slope of $1.0079{\pm}0.0135$, and an intercept of $-22\pm28$\,ppm, with a tight relation (not illustrated).  The scatter from that relation is virtually the same (close to 220 ppm) in each coordinate, suggesting that the two sets of eclipse depths have approximately the same uniformity.  We conclude that our procedure of choosing among four alternate sets of photometry does not degrade the uniformity of our results, but neither does it improve it significantly.  Given that different data sets can have potentially very different characteristics, we consider it prudent to use our adopted depths in our analyses reported below, but we also check the results using the GC depths.  Finally, we also have a third set of eclipse depths, obtained as the centroid of the distribution for eclipse depth, rather than the specific value selected using our Allan deviation slope criterion.  Those centroid-of-the-distribution (CD) depths are very close to our adopted values, as can be seen by comparing the vertical lines to the distributions on the top left panels of Figure~\ref{pix_coeffs_76} and Figure~\ref{pix_coeffs_131}.

\begin{figure}[htp]
\plotone{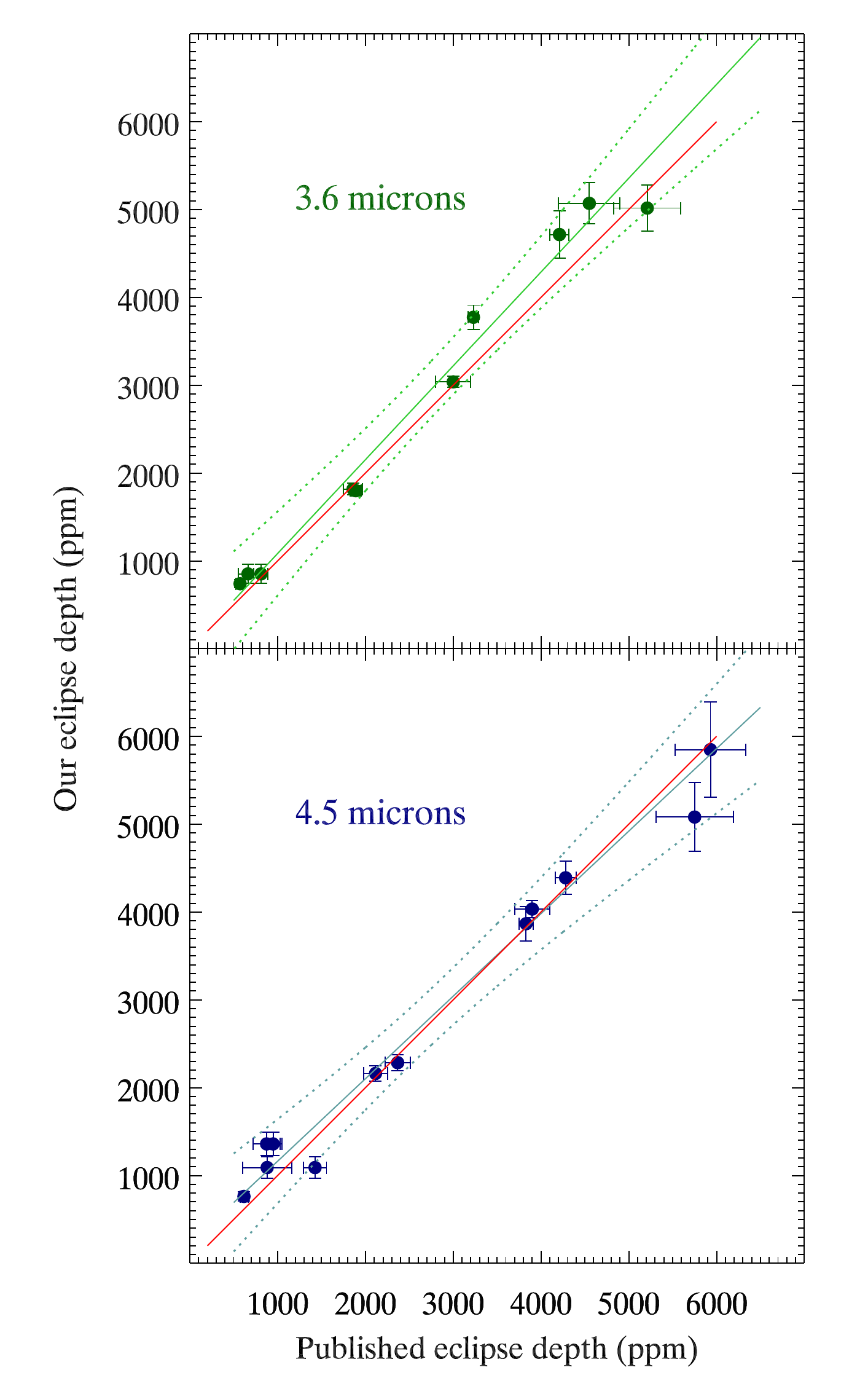}
\caption{Comparison between our derived eclipse depths and previously published results, for seven planets at 3.6\,$\mu$m, and adding WASP-62b at 4.5\,$\mu$m.  The red lines represent slopes of unity, i.e. perfect agreement.  The solid lines in the same color as the points are the results of maximum likelihood regressions, considering errors in both coordinates. The regression lines are of the form: $y-\overline{y}=a(x-\overline{x})+b$, where $\overline{x}$ is the average value.  The regression values of $a$ and $b$ are $1.068\pm0.081$ and $22\pm120$ at 3.6\,$\mu$m.  The 4.5\,$\mu$m regression values are $0.945\pm0.066$ and $25\pm101$.   The dotted lines are the $\pm3{\sigma}$ limits on departures from the regression lines.}
\label{others_fig}
\end{figure}

Finally, we examine how our eclipse depths correlate with values published in the peer-reviewed literature.  We make this comparison for seven planets at 3.6\,$\mu$m and eight planets at 4.5\,$\mu$m.  These planets and their previous eclipses are: HAT-P-13b observed by both \citet{hardy17} and \citet{buhler16}, KELT-2b \citep{piskorz18}, WASP-12b \citep{stevenson14}, WASP-14b \citep{wong15}, WASP-19b \citep{wong16}, and WASP-43b \citep{stevenson17}.  At 4.5\,$\mu$m, we added WASP-62b \citep{kilpatrick17}. Details of our comparisons for some of these cases are discussed under the notes for individual planets in the Appendix.  Although we have analyzed WASP-103b \citep{kreidberg18}, we omit it from our comparison, for the reason discussed in the notes for that planet.

Figure~\ref{others_fig} shows the comparisons between our eclipse depths and published values at both wavelengths.  Taking the published values as the independent variable ($x$), and our values as the dependent variable ($y$), we calculate the slope and zero-point of a linear relation, using the maximum likelihood regression method described by \citet{kelly07}, and accounting for errors in both $x$ and $y$.  The solutions also yield the standard deviation of slope and intercept. A main result of this paper is a systematic trend in exoplanetary brightness temperatures as a function of equilibrium temperature (Section~\ref{subsec_deviations}).  Since planets with the highest equilibrium temperatures tend to have the greatest eclipse depths, we want to verify that our main result will not be contaminated by a systematic error that trends with eclipse depth.  Comparing to previously published results, we expect to find slopes near unity, and small intercepts. The maximum likelihood regressions yield slopes and intercepts that are in $1-\sigma$ agreement with unity and zero, respectively - see the caption of Figure~\ref{others_fig}. We conclude that our eclipse depths do not deviate systematically from previous work.

\subsection{Dilution Corrections} \label{ssec_dilution}

Our photometry is normalized to unity during eclipse. When a stellar companion is present, that normalization can include contaminating light from the companion, thus requiring a dilution correction applied to the measured eclipse depths.  We identify systems needing dilution correction by inspecting the \textit{Spitzer} images themselves, and by consulting results from high resolution imaging \citep{ngo15, ngo16, wollert15, wollert_brandner, evans18}.

For systems with identified companions, we multiply our fitted eclipse depths times a dilution correction factor $f_d$ given as:
\begin{equation}
f_d = 1 + f_s r_s,
\end{equation}
where $f_s$ is the fraction of the light from the companion star that is scattered or diffracted into the photometric aperture centered on the target star (or completely in the aperture in some cases), and $r_s$ is the ratio of the total brightness of the companion star to the total brightness of the target star in a given \textit{Spitzer} band. Multiplying our fitted eclipse depth times $f_d$ yields the true astrophysical eclipse depth.  Twelve of the systems we analyze have stellar companions that are sufficiently bright and close that $f_d$ significantly exceeds unity.  Those twelve systems are listed in Table~\ref{dilute_table}, with our calculated $f_d$ factors. 

The twelve systems listed in Table~\ref{dilute_table} can be divided into two groups.  First, there are WASP-12, -49, -76, -103, HAT-P-33, and KELT-2, whose stellar companions are entirely contained in the photometric aperture used for our \textit{Spitzer} photometry ($f_s = 1$).  The remainder of the Table~\ref{dilute_table} systems have companions that contribute only a fraction of their light to our photometric aperture ($f_s < 1$).  

\begin{figure}[htp]
\epsscale{1.15}
\plotone{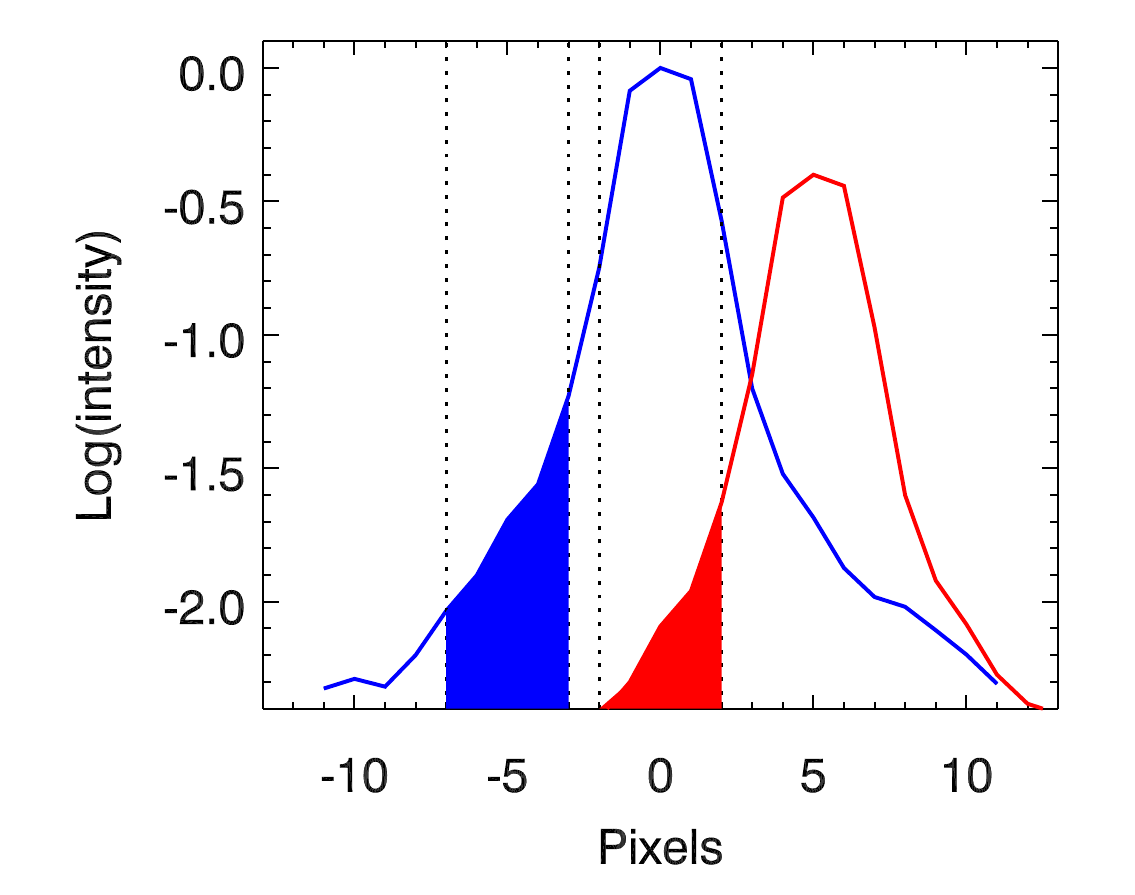}
\caption{Illustration of our method to define the effect of scattered or diffracted light from stellar companions lying outside of the photometric aperture (illustrated here in 1-D).  The bounds of the photometric aperture are shown as dashed lines at $\pm2$-pixels.  The red curve is the companion star, several times fainter than the target star (blue curve). The red and blue curves have the same shapes (PSFs), given their close proximity in the focal plane. The companion star light scattered into the photometric aperture is shaded in red.  We place a second aperture symmetrically located opposite to the companion star (dashed lines centered on $-5$-pixels), and we measure the light from the target star scattered into this symmetric aperture (shaded blue region).  Since the companion star contributes negligibly to the symmetric aperture, the ratio of the shaded blue region to the peak (or integral) of the blue curve equals the corresponding ratio for the red region and curve. The peak intensities of each curve (and integrals) are readily measured for each image, but we use average values for the entire time series.  We thereby solve for the amount of companion light in the shaded red region of each exposure without requiring complex manipulations of the data, and without needing to assume an analytic expression for the wings of the PSFs.}
\label{dilution}
\end{figure}

For this second group, we determined $f_s$ by placing an aperture at a position adjacent to the target star, choosing the location to be symmetrically opposite the contaminating star.  For example, if the contaminating star is 4 pixels below the target star, we place our aperture 4 pixels above the target star.  Our assumption is that the point-spread-function for the target star and the companion are the same, because they are both very close to the center of \textit{Spitzer's} field of view.  In that case, the fraction of target light scattered or diffracted into our symmetric aperture will be the same as the fraction of companion light scattered or diffracted into the target aperture. Also, the symmetric aperture is sufficiently distant from the companion star to be unaffected by light from the companion. We choose the symmetric aperture to have the same size as the target aperture. Figure~\ref{dilution} illustrates this method. For cases where we use a variable-radius aperture on the target star, we use a symmetric aperture having a constant radius closest in size to the median value of the variable aperture used for the target.

From the time series photometry, we determine the median value of the flux in the symmetric aperture, after subtracting a background value, and we divide that by the median background-subtracted flux measured for the target star, and the ratio of those fluxes is $f_s$. In the cases where the companion star is spatially separated from the target in the \textit{Spitzer} images, we calculate $r_s$ by fitting 2-D Gaussian functions to both stars, and calculate $r_s$ as the ratio of the areas under those Gaussians.  

The procedure described above does not require independent measurements of the spectral type or magnitude difference between the target and companion star.  Instead, we measure $r_s$ directly from the \textit{Spitzer} data.  However, for WASP-12, -49, -76, -103, HAT-P-33, and KELT-2, the companion stars are too blended with the target to make that direct measurement, and for HAT-P-30 the blend is also problematic.   In those cases, we estimate $r_s$ in the \textit{Spitzer} bands based on the difference in K-magnitudes, and the spectral types (effective temperatures) given by various sources (see the Appendix).   From those magnitudes and effective temperatures, we calculate the flux ratio in the \textit{Spitzer} bandpasses by interpolating among values output by the STAR-PET\footnote{http://ssc.spitzer.caltech.edu/warmmission/propkit/pet/starpet/} online calculator.

In addition to the correction factors listed in Table~\ref{dilute_table}, WASP-49 and WASP-121 have other stars at 9 and 7 arcsec distant, respectively, \citep{lendl12, delrez16},   Those companions are too faint and too distant in sky separation to significantly contaminate our \textit{Spitzer} observations, and no dilution correction is required.

Our dilution correction factors listed in Table-\ref{dilute_table} have not been applied to the 'as-measured' eclipse depths listed in Table~\ref{eclipse_depths}.  However, they have been applied before we use the Table~\ref{eclipse_depths} values in our subsequent analyses.

\section{Results for Orbital Phase}  \label{sec_phase}

Previous secondary eclipse observations have shown that the majority of transiting hot Jupiters have orbital eccentricities close to zero due to tidal circularization (e.g., \citealp{baskin13,  todorov13, beatty14, deming15, garhart18}).  

\begin{figure}[htp]
\epsscale{1.15}
\plotone{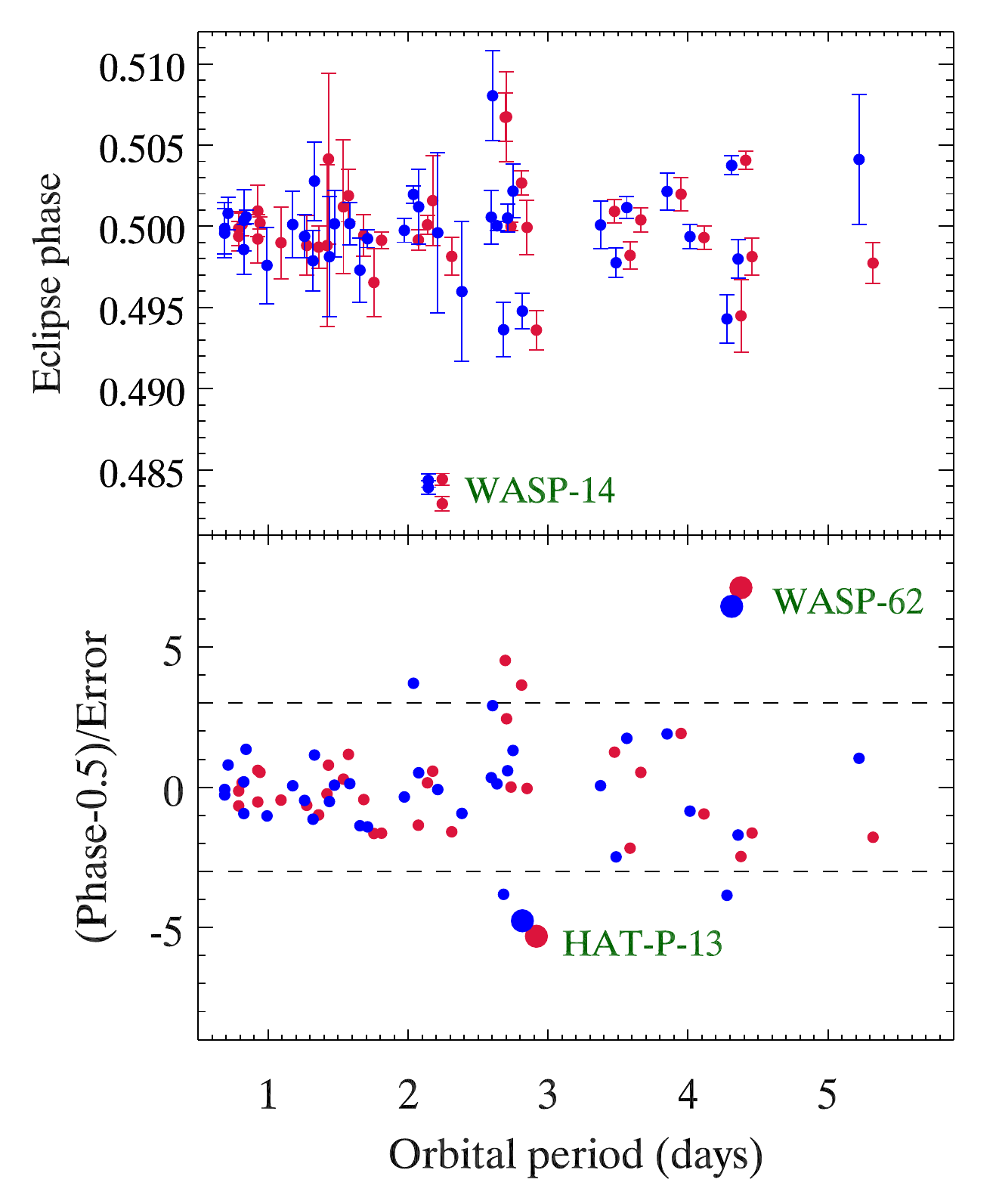}
\caption{The top panel plots the measured orbital phase of our secondary eclipses, corrected for light travel time across the orbit, versus the orbital period of each planet. Error bars include our measurement error and also imprecision in the orbital ephemerides. Red points are 3.6\,$\mu$m and blue are 4.5\,$\mu$m.  WASP-14b (labeled) is known to have an eccentric orbit, so the central phase deviates from 0.5. The lower panel plots the phase minus 0.5, divided by the error on the phase.  The scale of the ordinate is expanded, so WASP-14 is off-scale, and HAT-P-13 and WASP-62 are labeled (larger points).  Note the prominent deviation of WASP-62b due to a slightly eccentric orbit. Horizontal dashed lines mark $\pm3\sigma$.}
\label{phase_plot1}
\end{figure}

Our results are consistent with that trend.  The times and orbital phases of our observed eclipses are listed in Table~\ref{phase_table}.  The top panel of Figure~\ref{phase_plot1} shows our measured central phase for all of the eclipses we measure, corrected for light travel time across the orbit (a small effect, about $0.0002$ in phase), and plotted versus the orbital period of the planet.  For all planets, we add the precision of their orbital ephemerides in quadrature with the observed phase error to produce the error bars for phase on the figure. Two planets on Figure~\ref{phase_plot1} are already known to have eccentric orbits: WASP-14b \citep{blecic13, wong15}, and HAT-P-13 \citep{buhler16, hardy17}.  WASP-14b is labeled on the top panel of the figure.  

The bottom panel of Figure~\ref{phase_plot1} plots the deviation from phase 0.5 divided by the precision of the measurement (including ephemeris error), again versus the orbital period.  The scale of the ordinate is expanded, so that WASP-14b is now beyond the limits of the plot.  HAT-P-13b is labeled on this bottom panel, and also WASP-62b is labeled and has a clearly detected orbital eccentricity.  \textit{Spitzer} eclipse phases for WASP-62b agree very well between the two independent measurements, and the high statistical significance of the deviations ($>6\sigma$) makes the planet very obvious on the bottom panel of Figure~\ref{phase_plot1}.  The two measured phase values, corrected for light travel time are $0.50406\pm{0.00057}$ and $0.50375\pm{0.00058}$ at 3.6- and 4.5\,$\mu$m respectively.  The quoted errors again include imprecision in our improved ephemeris.  Weighting the phase in each band by the inverse of its variance yields an average orbital phase of $0.50391\pm{0.00041}$; the corresponding value of $e\cos{\omega}$ is $0.00614\pm{0.00064}$.  The orbital eccentricity of this planet is especially important because it is in the continuous viewing zone for JWST.  The eclipse occurs about 23 minutes later than phase 0.5, and that could potentially cause a significant degradation in JWST spectroscopy if the eclipse were incorrectly assumed to occur exactly at phase 0.5. 

\begin{figure}[htp]
\epsscale{1.15}
\plotone{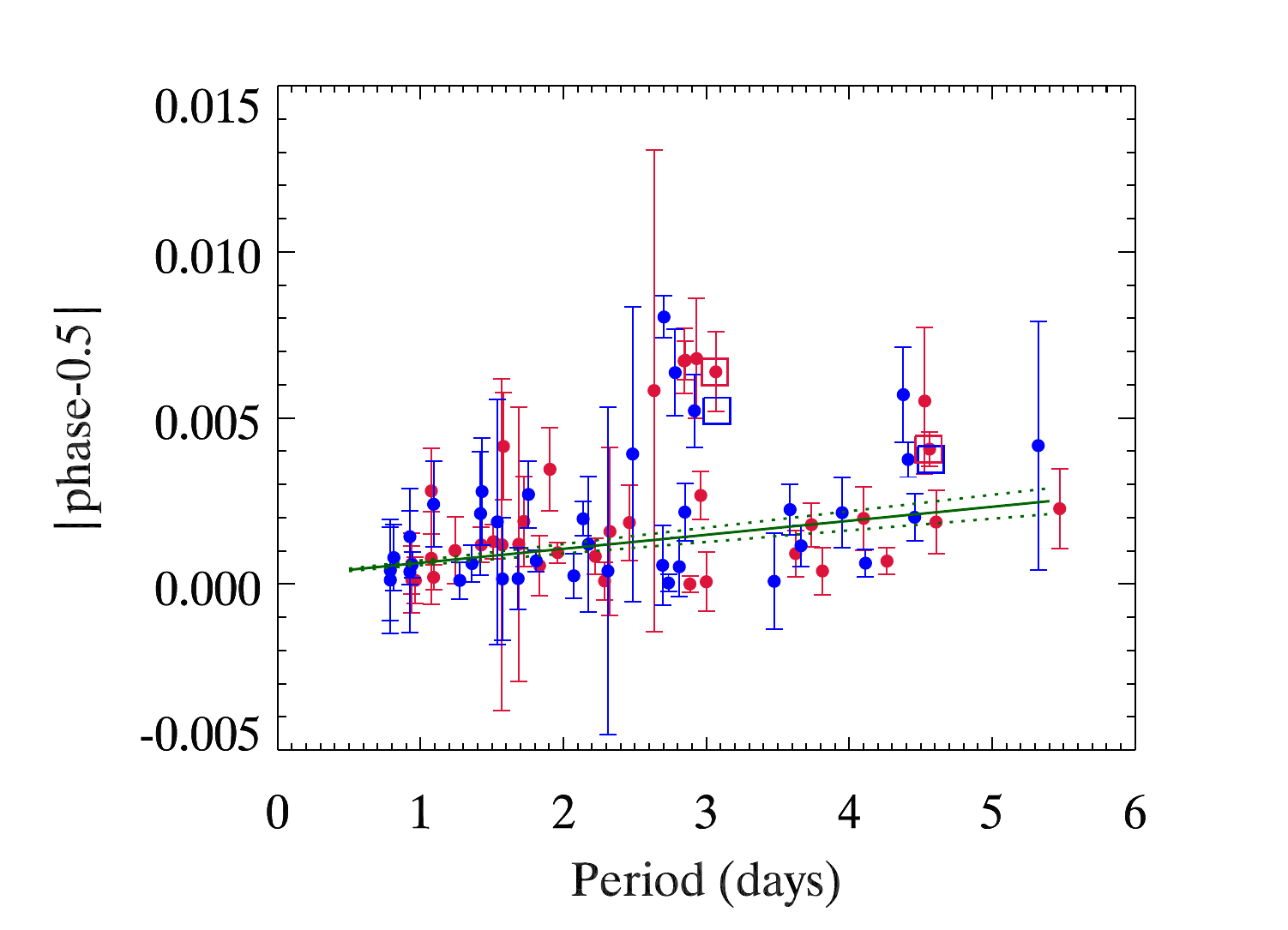}
\caption{Absolute deviation of the secondary eclipse phase from 0.5, versus orbital period in days.  The green line is a least-squares fit, ignoring WASP-14b (that is off scale). Dotted lines indicate the $\pm{1}\sigma$ error on the slope.  Planets with well established eccentric orbits (HAT-13b and WASP-62b) are plotted with open squares to distinguish them, and without error bars to minimize confusion.}
\label{abs_phase_plot}
\end{figure}

We have investigated whether the secondary eclipse phase deviates systematically from phase 0.5 at longer orbital periods, due to incomplete tidal circularization at greater orbital distances.  Figure~\ref{abs_phase_plot} shows the absolute deviation of the eclipse phase from 0.5, versus orbital period.  A least-squares fit accounting for the errors in phase yields a slope of $0.00043\pm{0.000072}$, if we ignore WASP-14b that would otherwise dominate the fit.  On that basis, the eclipse phase (on average) deviates from 0.5 by 0.00043 for each 1-day increase in orbital period.  If we also ignore WASP-62b and HAT-P-13b, the fitted slope becomes $0.00024\pm{0.000078}$.  However, those three planets are unambiguous examples of eccentric orbits, so ignoring them is ignoring the effect that we seek.  The fitted slope being $3\sigma$ above zero even when the obvious eccentric planets are ignored, is evidence for a lack of tidal circularization, increasing with orbital period in the range of our sample (0.8 to 5.5 days). However, the significance of the slopes depends on whether the distribution of central phases is Gaussian.  Ignoring the clearly eccentric planets, we applied an Anderson-Darling test to the distribution of central phases. This indicates a lack of normality (p-value of $10^{-4}$). That occurs because there are outliers, even after eliminating the clearly eccentric planets. Small amounts of orbital eccentricity in our sample could cause those outliers, and thereby contribute to failing the Anderson-Darling test. Moreover, the central portion of the phase distribution (within $\pm{2}\sigma$, comprising the majority of the phases) does pass the Anderson-Darling test, as described below. Hence we tentatively conclude that there is valid evidence for eccentricity increasing with orbital period.  However, eclipse phases are also sensitive to imprecision in orbital ephemerides, so this issue should be re-visited when more precise transit times and orbital periods become available (i.e., adding TESS data).

\begin{figure}[htp]
\plotone{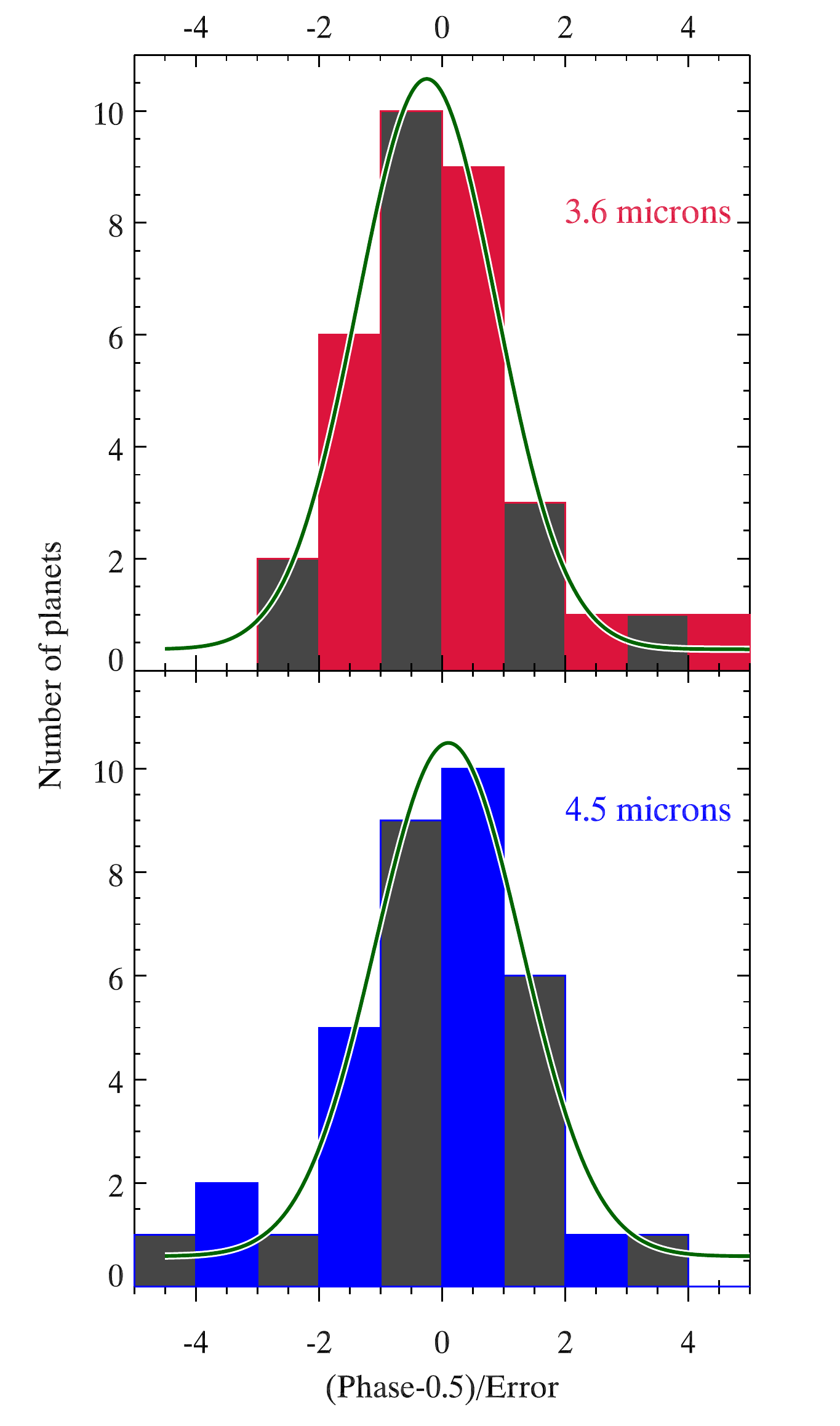}
\caption{Histograms of the deviation of our measured phase from 0.5, normalized by the error bar (i.e, histograms of the points on the lower panel of Figure~\ref{phase_plot1}. Alternate colors for adjacent histogram bins are used solely for visual clarity. We omit eccentric planets (WASP-14, WASP-62 and HAT-P-13).  The green curves are best-fit Gaussians (see text).}
\label{phase_plot2}
\end{figure}

Figure~\ref{phase_plot2} shows distributions of the phase offset from 0.5 for most planets, normalized by the error of each measurement, i.e., a histogram of the values plotted in the lower panel of Figure~\ref{phase_plot1}.  When constructing the histograms, we omitted WASP-14, WASP-62 and HAT-P-13, so the histograms represent only planets whose potential orbital eccentricity is not detected.  The green curves are the result of fitting Gaussian functions to the distributions defined by these histograms.  Fitting Gaussians to these binned distributions is a good way of measuring the dispersion in the central portion of the distribution, with minimal sensitivity to outliers. As noted above, the total phase distributions are not Gaussian, because they fail an Anderson-Darling test.  That failure occurs because there are more outliers than expected for a Gaussian distribution.  Repeating the Anderson-Darling test for phases within $\pm2$ standard deviations of 0.5 shows good normality (p-values of 0.60 and 0.75 at 3.6- and 4.5\,$\mu$m, respectively). If all planets represented in the central portions of the distributions have tidally circularized orbits with zero eccentricity, and if our errors are correctly estimated, then the fitted Gaussians should be centered at zero, with standard deviations of unity.  The fitted Gaussian functions come close to that expectation, but differ slightly.  The standard deviations of the Gaussians at 3.6- and 4.5\,$\mu$m are 1.13 and 1.19, respectively.  Given that those values exceed unity in both \textit{Spitzer} bands, and given the evidence discussed above for eccentricity increasing with orbital period, we conclude that there may be a small amount of undetected orbital eccentricity in our sample of planets. We emphasize that this conclusion is tentative and should be re-visited when more eclipses are analyzed, especially using improved ephemerides from TESS.

We are also interested in whether the average phase deviates from 0.5 systematically in one direction, such as the "uniform time offset" effect described by \citet{williams06}.  Although the binned histograms in Figure~\ref{phase_plot2} are good visual representations, and a good way of evaluating the scatter in the data compared to our estimated errors, they are not optimum for measuring potential systematic displacement.  The binning process slightly distorts the distributions \citep{kipping10}, and they effectively weight each measured phase by the inverse of its standard deviation, whereas correct weighting is proportional to the inverse of the variance (variance = standard deviation squared).  So we also use the original phase data (top panel of Figure~\ref{phase_plot1}), and we compute the average phase, correcting for light travel time and weighting each measurement by the inverse of its variance. We again omit WASP-14, WASP-62, and HAT-P-13.  We find average eclipse phases of $0.499969\pm{0.000157}$ and $0.500355\pm{0.000176}$ at 3.6- and 4.5\,$\mu$m, respectively.  If we combine the bands, we derive a grand average phase of $0.500140\pm{0.000118}$. Note that even with slightly non-zero eccentricities, the average phase should indeed be very close to 0.5, because $\omega$ is effectively random.  

Only with the uniform time offset effect described by \citet{williams06} would we expect to detect an average difference from phase 0.5. However, we find no statistically significant difference. Considering the average orbital period of our planet sample ($\sim 2.3$ days), our precision on the grand average phase corresponds to about 23 seconds.  That is comparable to the uniform time offset values calculated by \citet{williams06}, and eliminates some of their largest modeled offsets.  Our precision for this aggregate sample of planets is only modestly poorer than the offset actually detected (33~seconds) for the high signal-to-noise planet HD\,189733b by \citet{agol10}.  With a larger sample of secondary eclipses (by a factor of $\sim 4$), and with better ephemerides (less ephemeris error), it is reasonable to project that the average time offset value would be measurable using \textit{Spitzer} eclipses in a more extensive statistical study. 

\section{Converting Eclipse Depths to Brightness Temperature} \label{sec_bright}

The depth of a secondary eclipse is the ratio of flux from the planet to the flux from the star.  We convert eclipse depths to a brightness temperature for the planet's emission in both \textit{Spitzer} bands.  Before doing this, we correct the `as observed' depths (Table~\ref{eclipse_depths}) for dilution by companion stars using the factors in Table~\ref{dilute_table}.   We then divide the corrected eclipse depth by the ratio of solid angles (planet-to-star, based on their radii).  That quotient is the disk-averaged intensity of an equivalent blackbody for the planet, divided by the disk-averaged intensity of the star.  We represent the host stars using ATLAS model atmospheres \citep{kurucz}, rounding the stellar surface gravity to the nearest 0.5 in $\log (g)$, but interpolating in the model grid to the exact stellar temperature (usually as reported in the discovery paper of each planet). For both planet and star, we must account for the \textit{Spitzer} bandpass functions.  We multiply those functions times the stellar-disk-averaged intensity from the ATLAS models, and integrate over wavelength.  We do the same for a series of Planck functions whose temperatures bracket the temperature of the planet, and take the ratio to the bandpass-integrated stellar spectrum.  We then interpolate in that grid of bandpass-integrated intensity ratios to find the equivalent blackbody temperature that matches the ratio calculated from the eclipse depth.  That temperature is the brightness temperature of the planet in that particular \textit{Spitzer} band.  As for error bars, the precision of the planetary brightness temperature is dominated by the fractional error in the eclipse depth, so we propagate the eclipse depths error bars to the brightness temperatures.  Our observed brightness temperatures and errors are listed in Table~\ref{temp_table}, together with equilibrium temperatures for the planets.

In addition to the observed planets, we also calculate brightness temperatures for models of the planets (see Section~\ref{sec_spectra}).  We multiply the modeled spectra over the \textit{Spitzer} bandpass functions, integrate over wavelength, and interpolate in a grid of blackbodies, just as for the observed planets.  We also check the calculation by replacing the planetary modeled spectra with blackbodies, and verifying that the retrieved brightness temperature closely equals the temperature of the blackbody substitute (difference less than 1 Kelvin).  

\section{Implications for Heat Re-distribution} \label{sec_heat}

Secondary eclipses can be used to make statistical inferences concerning longitudinal heat redistribution on hot Jupiters \citep{cowan}. Figure~2 of \citet{cowan} shows that the average brightness temperature in our two \textit{Spitzer} bands should be a good approximation to the day side effective temperature. Therefore, given a value for the Bond albedo, redistribution of heat from stellar irradiance determines the day-side temperature, that can be inferred from the \textit{Spitzer} eclipse depth. The hottest planets tend to have low albedos because they are too hot for significant cloud condensation \citep{sudarsky00}.  To the extent that their albedos approach zero, their eclipse depths are therefore indicative of the degree of longitudinal heat redistribution. Although infrared phase curve observations are the gold standard for measuring longitudinal heat redistribution, it is easier to observe a large sample of infrared eclipses than the same number of phase curves. Hence, eclipses can usefully speak to the statistical properties of heat redistribution, especially in the strong irradiance limit.  We calculate the observed day side temperature for each planet in our sample, assumed to equal an average of the 3.6- and 4.5\,$\mu$m brightness temperatures, weighted by the inverse square of their errors.  (For planets without 3.6\,$\mu$m eclipses, we use the 4.5\,$\mu$m brightness temperature.)

\begin{figure}[htp!]
\epsscale{1.15}
\plotone{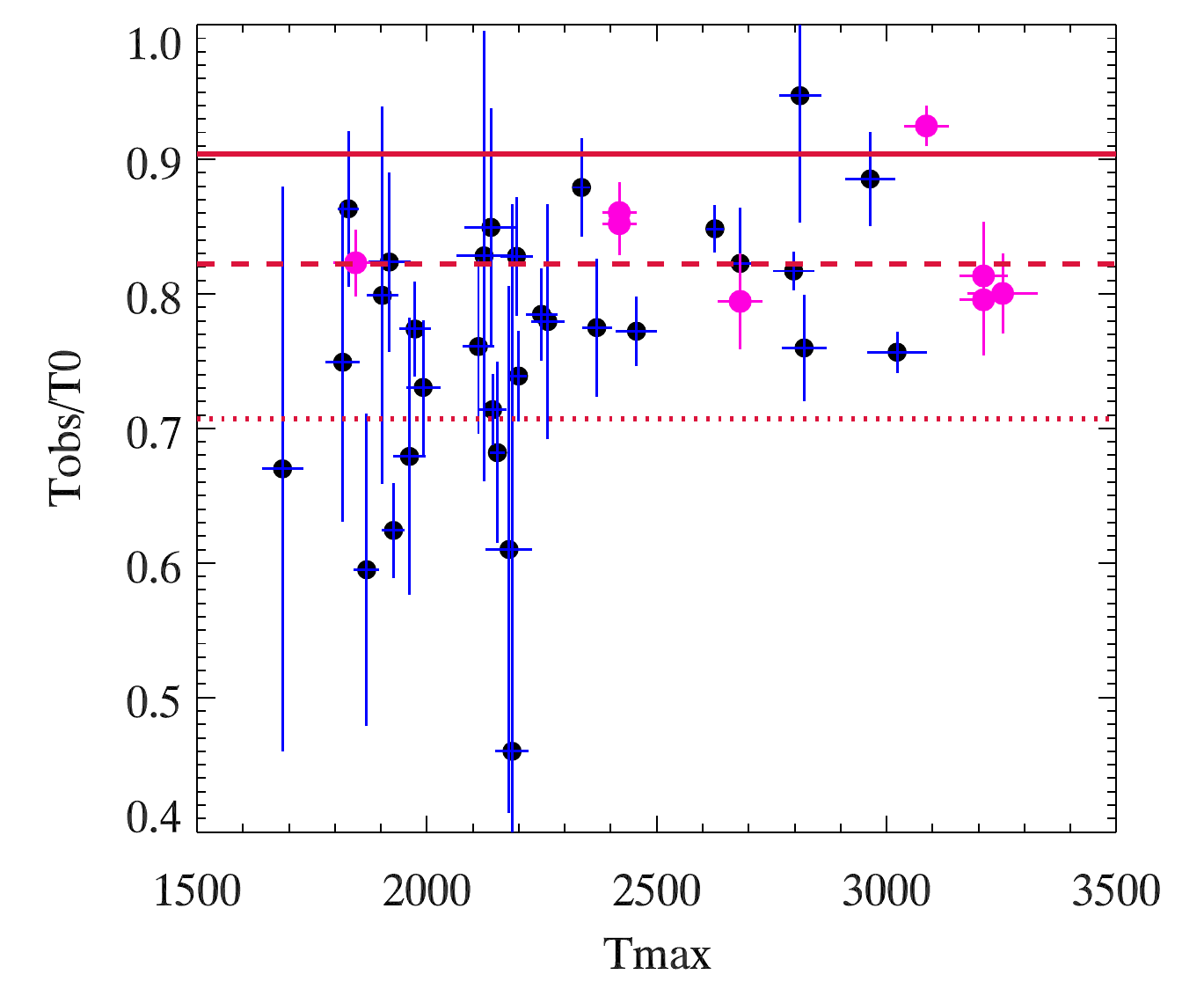}
        \caption{Figure~7 of \citet{cowan} replotted with the 36 hot Jupiters analyzed in this paper. WASP-14b, -19b, and -103b are represented by two eclipses each (as per Table~\ref{eclipse_depths}), so there are 39 total points. $T_{obs}$ is our observed day side temperature, from the error-weighted average of the 3.6- and 4.5\,$\mu$m brightness temperatures.  $T_0$ is the equilibrium temperature that would result at the sub-stellar point with zero albedo, in the limit of no redistribution of heat. $T_{max}$ is the calculated day side equilibrium temperature with zero albedo and no redistribution of heat. Just as in their original figure, the solid line shows zero recirculation, the dashed line is a uniform day-hemisphere, and the dotted line is a uniform planet. An albedo of zero was used to calculate the red lines. Planets with published \textit{Spitzer} phase curves are plotted in magenta, but using values from our eclipse results.} 
\label{fig_heat}        
\end{figure}

\begin{figure}[ht!]
\plotone{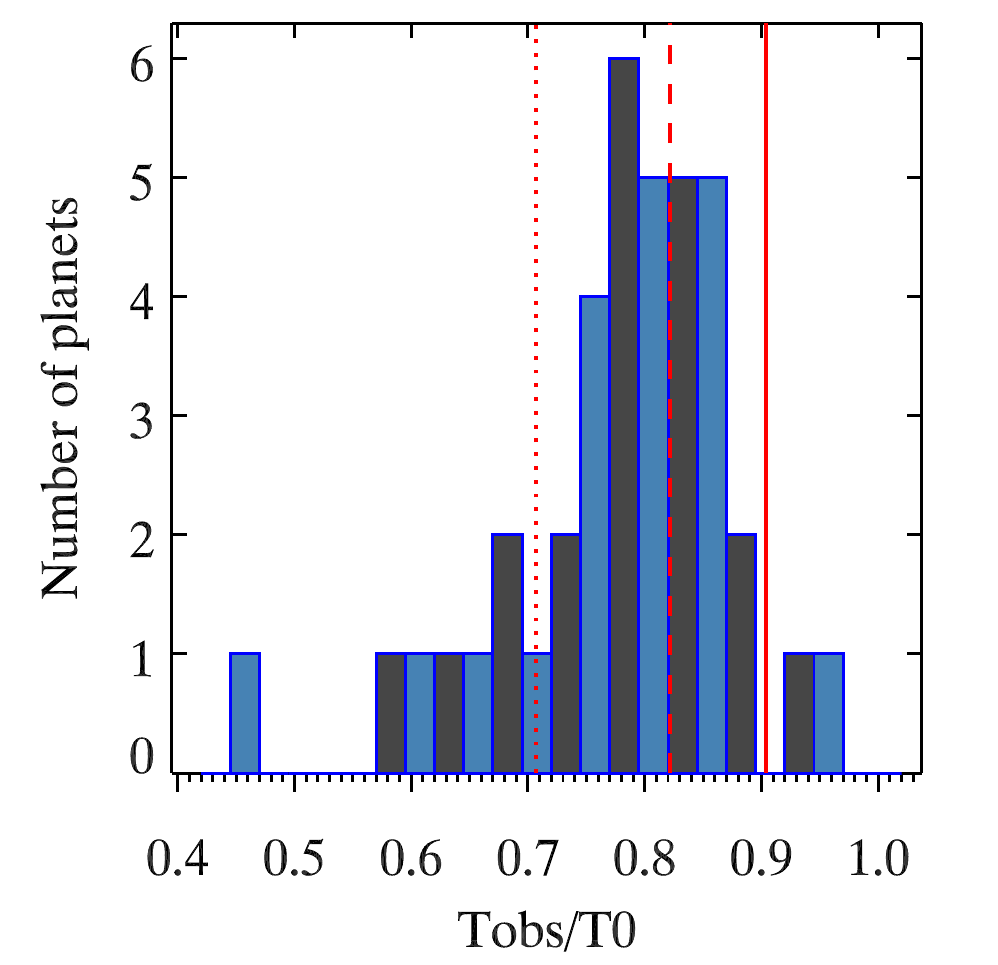}
        \caption{Histogram of $T_{obs}/{T_0}$ values from the Y-axis of Figure~\ref{fig_heat}. $T_{obs}$ is our observed day side temperature, from the error-weighted average of the 3.6- and 4.5\,$\mu$m brightness temperatures.  $T_0$ is the equilibrium temperature that would result at the sub-stellar point with zero albedo, in the limit of no lateral redistribution of heat. Alternate colors for adjacent histogram bins are used solely for visual clarity. As in Figure~\ref{fig_heat}, the solid red line shows zero heat redistribution, the dashed line is a uniform day-hemisphere, and the dotted red line is a uniform planet. The median value of $T_{obs}/{T_0}$ is 0.79 for our sample, very close to uniform day-side hemispheres. An albedo of zero was used to calculate the red lines. Planets falling left of the dotted red line must have albedos significantly greater than zero.} 
\label{fig_heat_hist}        
\end{figure}

Figure~\ref{fig_heat} uses the observed day side temperatures for the 36 hot Jupiters analyzed here in a replication of Figure~7 from \citet{cowan}.  The X-axis is the {\it calculated} maximum day side temperature, assuming zero albedo and no redistribution.  The Y-axis is the {\it observed} day side temperature, normalized to the maximum equilibrium temperature at the sub-stellar point, as described by \citet{cowan}.  Our version of this figure has less scatter than the original from \citet{cowan}. (Although our sample is not identical to \citealp{cowan}, they did predict that reduced scatter would be possible with a uniform analysis.) Notice that no planet lies in the unphysical region above the solid line by more than $1.4\sigma$.  The figure suggests a division into two regimes.  The hottest planets ($T_{max}>2200$K) all lie above the dotted red line that indicates uniform redistribution. About 35\% of planets whose calculated maximum temperature falls between $\sim 1700$K and $\sim 2200$K require non-zero albedos (below the dotted red line), even if their redistribution of stellar irradiance is uniform over the entire planet.  We interpret this division as being due to a combination of factors, including the onset of cloud condensation at the cooler temperatures (increasing the albedo), as well as the hydrodynamic properties of the circulation, which inhibit efficient redistribution at the highest levels of irradiance \citep{komacek17, parmentier18a}.  The planets hotter than $T_{max}\sim 2200$K are distributed near the dashed red line corresponding to zero albedo and uniform redistribution only on the day-side hemisphere. While some of these planets may have Bond albedos significantly exceeding zero (e.g., WASP-12b, \citealp{schwartz17}), our eclipse data do not require that because we do not find any of the hottest planets lying below the dotted line on Figure~\ref{fig_heat}.  Figure~\ref{fig_heat_hist} shows a histogram of the $T_{obs}/T_0$ values for all 36 planets, illustrating that the peak of the distribution is very close to the dashed line.  We note that common practice in the community is to estimate the temperature of hot Jupiters (e.g., in discovery papers) by adopting zero albedo and uniform redistribution.  Figure~\ref{fig_heat} shows that uniform day-side redistribution is more accurate for the hottest planets.

Six planets in our sample (WASP-12, -14, -18, -19, -43, and -103) have published \textit{Spitzer} phase curves. Those planets are plotted in magenta on Figure~\ref{fig_heat} (but using our eclipse results), and they are typical of the hotter group. Therefore we conclude that the \textit{Spitzer} phase curve results for the hottest planets represent an unbiased sample.   

\section{Implications for Emergent Spectra and Atmospheres}  \label{sec_spectra}

We now discuss the implications of our secondary eclipse depths for the emergent spectra of hot Jupiters, and for physical conditions in their atmospheres.  As prelude to the results, we first explain the rationale for a statistical approach (Sec.~\ref{subsec_stat}), and we describe two sets of modeled spectra that we use in this study (Sec.~\ref{subsec_models}).  Our results for the planets (Secs.~\ref{subsec_deviations} to ~\ref{subsec_select}) differ from expectations based on classic 1-D model atmospheres, and in Sec.~\ref{subsec_structure} we discuss that difference in terms of the atmospheric structure of the planets.

\subsection{A Statistical Approach} \label{subsec_stat}

\begin{figure}[ht!]
        \includegraphics[scale=0.45]{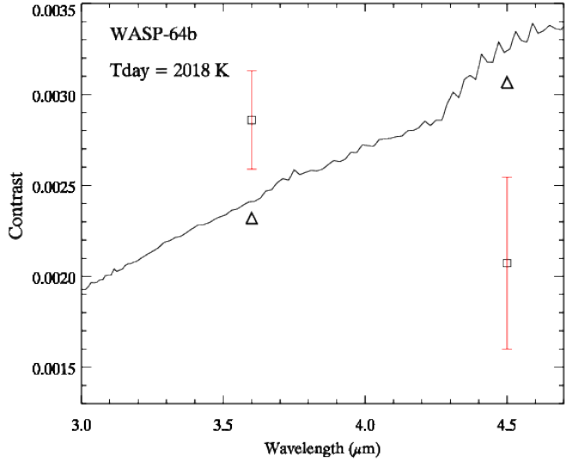}
        \advance\leftskip0.3cm
        \caption{Example of fitting 3.6- and 4.5\,$\mu$m eclipse depths to a blackbody planet, with an ATLAS model atmosphere to represent the star. The squares with error bars are the observed eclipse depths, and the triangles are the values expected from integrating models of the planet and star over the \textit{Spitzer} bandpasses, using the IRAC response functions.  WASP-64b has an equilibrium temperature of 1674K, assuming zero albedo and redistribution uniformly over both the day and night hemispheres. The best-fit blackbody has a temperature of 2018K, consistent with less than uniform redistribution (as per Figure~\ref{fig_heat}). The ripples in the best-fit curve are due to spectral structure in the modeled {\it stellar} spectrum.} 
\label{fig_2chan_example}        
\end{figure}

The earliest results for \textit{Spitzer's} secondary eclipses of hot Jupiters were interpreted in terms of molecular absorptions (e.g., \citealp{madhu11}).  \citet{hansen} questioned whether molecular features can be reliably detected using \textit{Spitzer's} photometry. Figure~\ref{fig_2chan_example} shows an example of fitting eclipse depths in those two channels to a blackbody planet.  This fit yields a good estimate for the day side temperature of the planet.  However, due to modest signal-to-noise and the lack of molecular band shape information, it is not typically possible to confidently associate molecular features with deviations from the best-fit blackbody.

Rather than attempting to identify molecular absorptions in individual planets, we adopt a statistical approach wherein we look for trends in our total sample.  Pioneering work of this type was reported by \citet{triaud14a, triaud14b, beatty14, beatty19}, and also \citet{kammer15}, \citet{adams18}, and \citet{wallack18}.  A statistical approach to {\it transit} (not eclipse) spectroscopy was elucidated by \citet{sing16}.  Our statistical approach differs somewhat from past work, as we explain in Sec.~\ref{subsec_deviations}.  

\subsection{Two Sets of Models}  \label{subsec_models}

\begin{figure}[ht!]
\epsscale{1.25}
\plotone{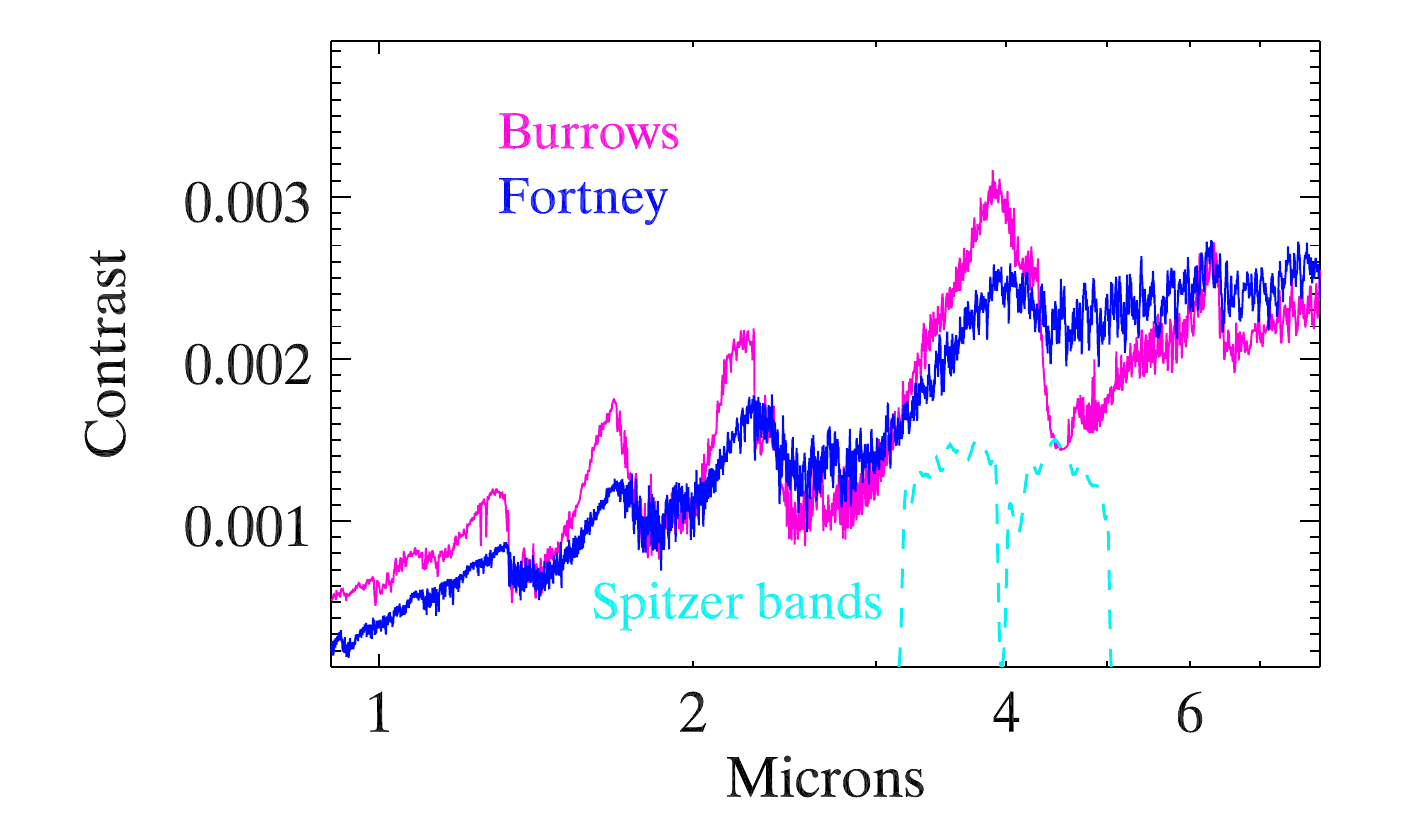}
        \caption{Comparison of Burrows and Fortney modeled spectra for planetary and stellar parameters equal to our average host star, and average planet.  Those parameters are $T_{s}=6040$K, $R_s=1.4R_{\odot}$,  $R_p=1.4R_{J}$, and $M_p=1.5M_{J}$.  The Burrows model lies at an orbital distance of 0.025~AU, versus 0.018~AU for the Fortney model.  Their day side temperatures are closely similar, due to different prescriptions for redistribution of stellar irradiance (see text, Sec.~\ref{subsec_models}).  The Burrows model has stronger spectral features, as discussed in Sec.~\ref{subsec_models}.  The \textit{Spitzer} band response functions at 3.6- and 4.5\,$\mu$m are included for comparison. }
\label{model_compare}        
\end{figure}

We use two sets of well documented model atmospheres for the planets, from Adam Burrows \citep{burrows97, burrows06} and Jonathan Fortney \citep{fortney05, fortney08}. Rather than calculating individual models for each of the 36 hot Jupiters in our sample, we model the planets using 'tracks' wherein the stellar insolation varies in magnitude.  We adopt stellar and planetary mass and radius based on the median values of our sample, thereby making an average hot Jupiter orbiting an average star. We vary the planetary temperature by placing that average planet at different orbital distances, and we use solar metallicity cloudless atmospheres for all models.  The Burrows and Fortney codes use different treatments of heat redistribution: Fortney adopts a uniform redistribution over both day and night hemispheres, whereas Burrows redistributes approximately over the day hemisphere, and partially into the night hemisphere. The consequence is that the Fortney models are cooler than the Burrows models at a given orbital distance.  But a Fortney model at an orbital distance of $a/\sqrt{2}$ should produce a comparable spectrum to a Burrows model orbiting at distance $a$; in particular it will have a very similar day side effective temperature (total energy re-radiated).  That comparison is shown in Figure~\ref{model_compare}.  

The two spectra in Figure~\ref{model_compare} indeed have close overall flux levels, and spectral features that correspond in relative strength and shape versus wavelength, but not in total amplitude.  The Burrows models have overall deeper absorption features than the Fortney models at the same effective temperature.  The reason for that difference is not obvious, due to the complexity of the models. A myriad of possible differences can come into play, and fully exploring the underlying physics is beyond the scope of this paper.   As one example, the different treatments of longitudinal heat redistribution can also affect the vertical temperature structure, and different temperature structures as a function of optical depth will produce different emergent spectra. Fortunately, our principal result is not affected by the differences between the two sets of models, as we discuss in Sec.~\ref{subsec_deviations}.  Also, we find that the two sets of models produce tracks that conveniently bracket the observed locus of the planets. We thereby use the models to gauge the average magnitude of absorption features in the exoplanetary spectra (Sec.~\ref{subsec_deviations}). 

We also utilize both Burrows and Fortney models that feature temperature inversions.  The Burrows inverted models were computed by adding extra absorbing opacity between 0.003 and 0.6 bars, and preserving flux-constancy.  The inverted Fortney models simply specified temperature to increase linearly with decreasing log of pressure below one bar ($dT/d\log{P} = -160$K).  Those models are not flux-constant, but we use them only to explore how the inverted profiles affect the relative brightness temperature of the planets in the two  \textit{Spitzer} bands (Sec.~\ref{subsec_structure}).

\subsection{Deviations from Blackbody Spectra} \label{subsec_deviations}

Several statistical treatments have examined \textit{Spitzer} colors of hot Jupiters versus their brightness in a particular band (i.e, an HR-diagram analogy).  That approach is particularly useful when the luminosity of the planet is produced by an internal source.  But hot Jupiters primarily re-radiate external energy from their star, and their emergent spectrum is determined to first order by the level of irradiance.  In this case we find it useful to relate the planetary brightness temperatures in the two \textit{Spitzer} bands, rather than to correlate color with total brightness.  In other words, we want to study the shape of the emergent spectrum, not the total luminosity of the planet.   

\begin{figure*}[t]
	\includegraphics[width=\textwidth]{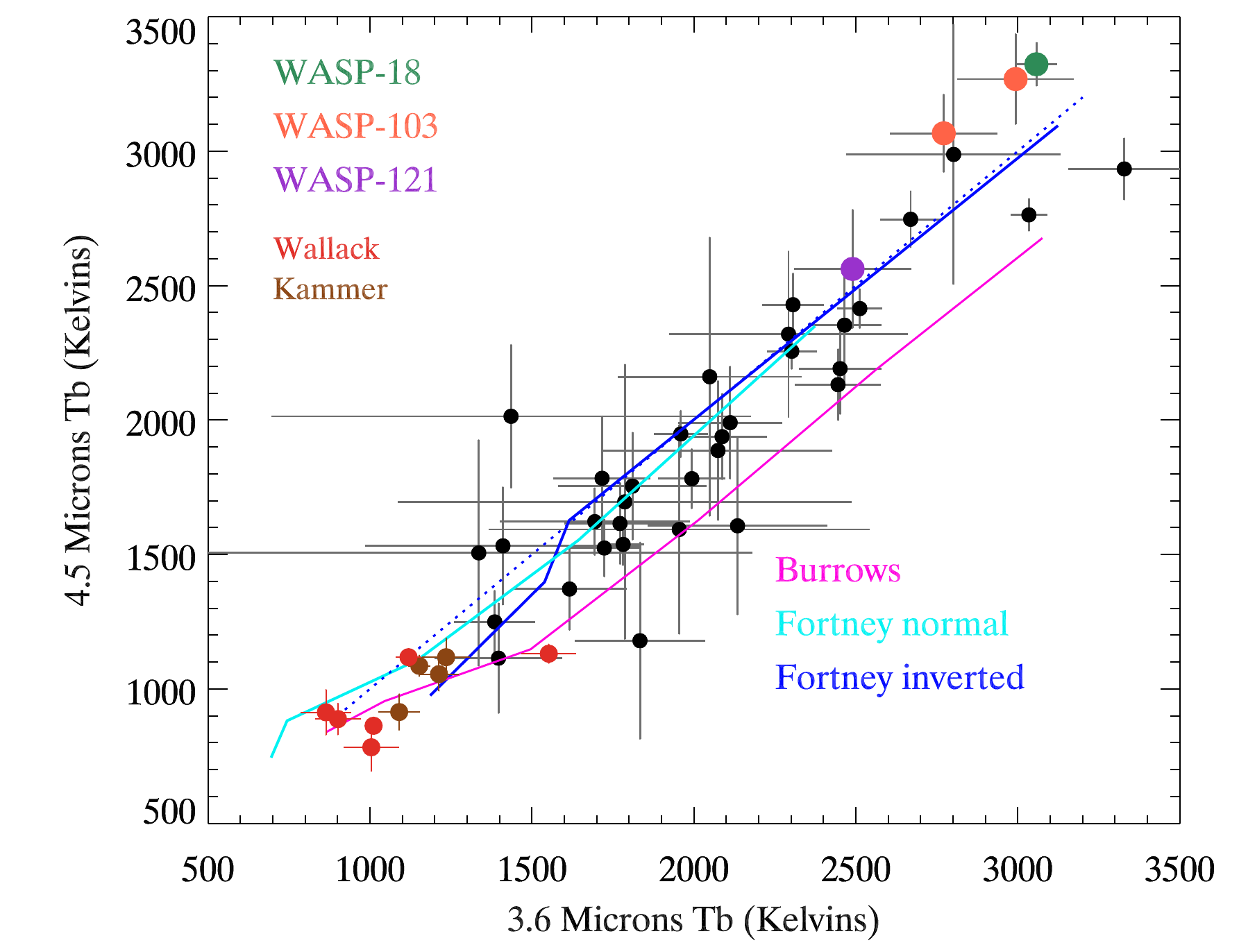}
	\caption{Comparison of brightness temperatures in both \textit{Spitzer} bands.  Blackbody planets would fall along the dotted blue line which has a slope of unity. Tracks of models from Burrows and Fortney are shown, and the inverted Fortney track merges with the normal track at high temperature (see text, Sec.~\ref{subsec_structure}), and the inverted Burrows track (not illustrated) does also.  The observations indicate that the planets are close to blackbodies, but cooler planets tend to have lower brightness temperatures at 4.5\,$\mu$m compared to 3.6\,$\mu$m, whereas hotter planets tend to be brighter at 4.5\,$\mu$m compared to 3.6\,$\mu$m.  Brightness temperatures from \citet{kammer15} and \citet{wallack18} are included in order to enhance the comparison for the coolest planets observed in secondary eclipse. The planets from \citet{kammer15} are HAT-19b, WASP-6b, -10b, and -39b.  The planets added from \citet{wallack18} are HAT-12b, -18b, -20b, and WASP-8b, -69b, and 80b.  See Sec.~\ref{subsec_deviations} and Sec.~\ref{subsec_slope} for discussion.}
\label{Tb_results}
\end{figure*}

Figure~\ref{Tb_results} shows the brightness temperature of our planets at 4.5\,$\mu$m versus their 3.6\,$\mu$m brightness temperature.  The model tracks from Burrows and Fortney are included, and the relation for purely blackbody planets ($T_{4.5} = T_{3.6}$) is shown as a dotted blue line.  In general, the Fortney model track passes through the envelope of the observed planets, and the Burrows track lies at the lower envelope (we quantify those statements below).  The 4.5\,$\mu$m \textit{Spitzer} band contains strong opacity from both water vapor and carbon monoxide, that is especially manifest in the Burrows spectra compared to Fortney (see Figure~\ref{model_compare}).  That causes the Burrows models to have a lower 4.5\,$\mu$m brightness temperature than Fortney, and thereby the Burrows track lies lower.  The 4.5\,$\mu$m band is thus indicative of overall stronger absorptions in the Burrows models versus Fortney (as per Figure~\ref{model_compare}), and we find that difference to be very useful as a diagnostic of the spectra of the planets. Most (60\%) of the observed planets lie between the two model tracks, indicating that the amplitudes of their spectral absorptions (especially at 4.5\,$\mu$m) are intermediate between the Burrows and Fortney models. That is an interesting inference, because to date there is little information on the magnitude of spectral features that applies to a comparably large sample of hot Jupiters. 

We quantify the differences between the observed planets and the model tracks by fitting a straight line to the 4.5- versus 3.6\,$\mu$m brightness temperature measurements, using the maximum likelihood method of \citet{kelly07}. From the precision on the intercept of that line, we find that the Fortney inverted track is $2.6\sigma$ higher than the average of the observations (not surprising, since temperature inversions have been found for only a few hot Jupiters).  The Fortney 'normal' models agree well with the observations (being only $0.04\sigma$ higher), and the Burrows track is $11\sigma$ below the observations, reflecting the strong modeled absorption in CO (Figure~\ref{model_compare}). The slope of the fitted line is greater than unity ($1.085\pm0.040$, $2.1\sigma$ above unity). Although that slightly steeper-than-unity slope is not statistically secure, it is suggestive, especially because there have been previous hints of that effect.
\citet{kammer15} and \citet{wallack18} found that cool Jupiters ($T < 1200$K) tend to have lower brightness temperatures at 4.5\,$\mu$m than at 3.6\,$\mu$m. \citet{beatty19} examined brightness temperatures in the two \textit{Spitzer} bands as a function of equilibrium temperature for hot Jupiters with phase curves, and their data suggest (but do not prove) a greater slope at 4.5- versus 3.6\,$\mu$m, consistent with our Figure~\ref{Tb_results}.  Beyond hot Jupiters, it has long been known that the exo-Neptune GJ\,436b ($T \sim 800$K) exhibits a puzzling flux excess at 3.6\,$\mu$m, that was attributed to disequilibrium chemistry \citep{stevenson10}, and a similar effect was recently found in GJ\,3470b \citep{benneke19}.   We hypothesize that Figure~\ref{Tb_results} hints at a pervasive and general effect that occurs over a large range of equilibrium temperature, and we investigate further using a physically somewhat different relation: the ratio of 4.5- to 3.6\,$\mu$m brightness temperature as a function of equilibrium temperature.

\subsection{A Slope in Brightness Temperaure Ratio}\label{subsec_slope}

\begin{figure*}[ht!]
        \includegraphics[width=\textwidth]{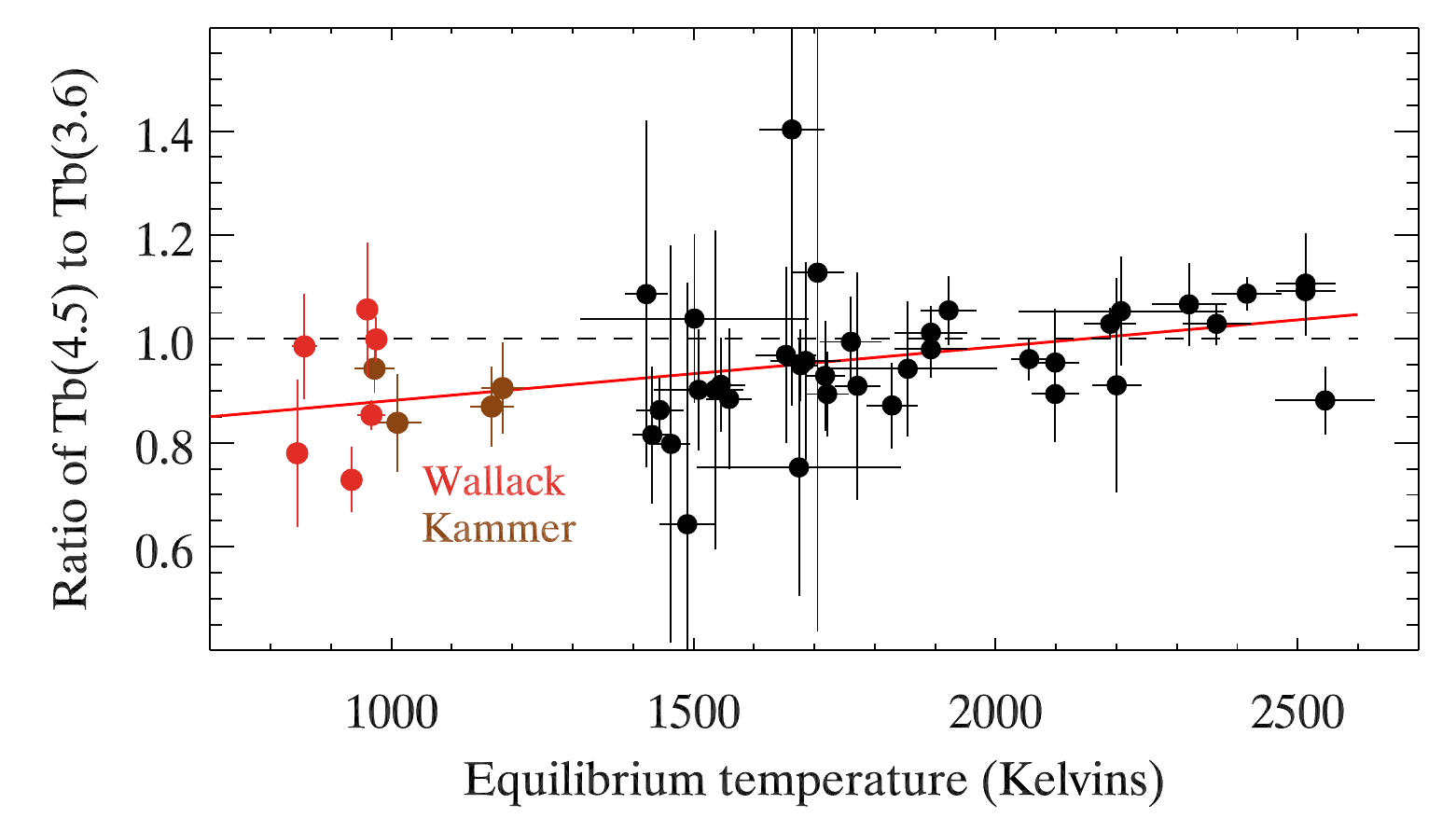}
        \caption{Ratio of the 4.5- to 3.6\,$\mu$m brightness temperature for our planets, plus planets from \citet{kammer15} and \citet{wallack18}.  The brightness temperature ratio is shown versus exoplanetary equilibrium temperature.  The ratio would be constant at unity for blackbody planets (dashed line), but a maximum likelihood regression \citep{kelly07} indicates an upward slope (red line), significant at $4.1\sigma$. } 
\label{slope_fit_ratio}        
\end{figure*}

If indeed the 4.5\,$\mu$m versus 3.6\,$\mu$m brightness temperature relation has a slope that exceeds unity, then the {\it ratio} of those brightness temperatures should be an increasing function of the equilibrium temperature of the planets, whereas the ratio would be constant (slope equal to zero) for blackbody planets.  The observed relation (including the planets from \citealp{kammer15} and \citealp{wallack18}) is shown in Figure~\ref{slope_fit_ratio}, and a maximum-likelihood regression yields a slope of $100\pm24$ parts-per-million (ppm) per Kelvin.  That slope is significant at $4.1\sigma$, and is obvious on Figure~\ref{slope_fit_ratio}. We confirmed the statistical significance using a nonparametric (Kendall-Tau) test.  Kendall-Tau rejects the null hypothesis of uncorrelated data with a p-value of 0.0012.  For each 1K increase in equilibrium temperature, the ratio of brightness temperatures (4.5 to 3.6) increases by 0.01\%.  Thus, from 800K to 2500K (for example), the ratio increases by 0.17, as shown by the red line on Figure~\ref{slope_fit_ratio}. 

We repeated the maximum-likelihood slope solution by omitting the planets from \citet{kammer15} and \citet{wallack18}.  That solution gives a slope of $152\pm49$ ppm per Kelvin, still significant at $3.1\sigma$.  We also repeated the solution using our set of GC eclipse depths (Section~\ref{ssec_checks}, but including \citealp{kammer15} and \citealp{wallack18}), and that decreases the slope to $87\pm26$\,ppm per Kelvin, still significant at $3.3\sigma$.  As a third possible case, we use our set of PD eclipse depths (also described in Section~\ref{ssec_checks}), and the slope is $105\pm24$ ppm per Kelvin, significant at $4.4\sigma$.  We also explored to what effect the significance of the result depends on the size of the per-point error bars.  Increasing the size of the per-point error bars by the (arbitrary, but implausible) factor of 1.5 (including the planets from \citet{kammer15} and \citealp{wallack18}), decreases the significance of the slope, but only to $3.1\sigma$.  Considering also that the Kendall-Tau test is independent of the error bars, we conclude that the observed planets robustly deviate from the blackbody line in the sense that hotter planets tend to become more prominent at 4.5\,$\mu$m relative to 3.6\,$\mu$m.  

A corollary of our conclusion is that the planets also robustly deviate from the model tracks, not merely from blackbodies. Specifically, the normal Fortney track has a slope of +19\,ppm per Kelvin (in Tb(4.5)/Tb(3.6) vs. $T_{eq}$) for $T>1000$K (versus the observed slope of $100\pm24$\,ppm per Kelvin), and the modeled ratio increases sharply to $>1.1$ for $T<1000$K, due to methane absorption in the 3.6\,$\mu$m band, very unlike the observations. 

We also investigated whether the Tb(4.5)/Tb(3.6) ratio correlates with stellar host temperature, and we find a $2.2\sigma$ effect.  However, planetary equilibrium temperature is a function of stellar temperature, so we would expect some degree of correlation with stellar temperature as a by-product of the correlation with planetary equilibrium temperature.  The stronger correlation of Tb(4.5)/Tb(3.6) with planetary equilibrium temperature indicates that the the temperature of the host star {\it per se} is not a primary factor.

\subsection{A Selection Effect?} \label{subsec_select}

We first consider whether the slope on Figure~\ref{slope_fit_ratio} could be due to a selection effect.  Eclipses in \textit{Spitzer's} 3.6\,$\mu$m band are harder to detect than at 4.5\,$\mu$m.  If the cooler planets have undetectable 3.6\,$\mu$m brightness temperatures, then the sample will tend to be incomplete for cool planets with high brightness temperature ratios (4.5 divided by 3.6).  That will bias the slope in the direction that we observe.  To evaluate whether this is a significant effect, we add five planets that are not currently included on Figure~\ref{slope_fit_ratio} because their eclipses were too weak to measure at 3.6\,$\mu$m. Those are WASP-75b and -49b (Figure~\ref{all36}), WASP-67b from \citet{kammer15}, and HAT-P-26b (and -17b at 4.5\,$\mu$m) from \citet{wallack18}.  For each of those planets, we postulate a 3.6\,$\mu$m eclipse depth that equals twice the error of the fit, a $2\sigma$ 'detection'.  Using a hypothetically minimal detection is conservative in this context, because it will maximize the brightness temperature ratio, while remaining consistent with the fact that the eclipses are not detected.  Adding those five planets, the significance of the slope on Figure~\ref{slope_fit_ratio} indeed decreases, but only from $4.1\sigma$ to $3.9\sigma$.  We conclude that a selection effect is not sufficiently strong to produce the slope that we observe, and we turn to possible astrophysical explanations.   

\subsection{Atmospheric Temperature Structure} \label{subsec_structure}

Since the emergent flux from exoplanetary atmospheres is directly related to the atmospheric source function (= the Planck function in LTE), it is virtually axiomatic that the slope we observe is related to the temperature structure of the atmospheres (i.e., temperature versus optical depth).  A prominent type of perturbation to exoplanetary atmospheric structure is the possible presence of temperature inversions.  Inversions have a long and popular history in exoplanetary science (e.g., \citealp{hubeny03, knutson08, knutson09, nymeyer11, haynes15, beatty17, sheppard17, arcangeli18, kreidberg18, mansfield18}).  \textit{Spitzer's} 4.5\,$\mu$m band is formed high in the atmosphere \citep{burrows07}, so an atmospheric temperature rising with height can in principle produce an excess brightness temperature at 4.5\,$\mu$m relative to 3.6\,$\mu$m.  Strong stellar irradiance provides the energy to maintain inversions, so a ratio of brightness temperatures (4.5 to 3.6) that increases with equilibrium temperature (as we observe) is at least qualitatively consistent with temperature inversions.  Nevertheless, we do not conclude that temperature inversions are the dominant effect that we are observing in Figure~\ref{slope_fit_ratio}.  Instead, we believe that the dominant effect is more subtle and pervasive than the temperature inversion phenomenon, as we now discuss.
 
Since \textit{Spitzer's} 4.5\,$\mu$m band contains both strong water vapor opacity, and the strong 1-0 band of carbon monoxide, it is indeed sensitive to high altitude temperature inversions.  Three planets in our sample (WASP-18b, -103b, and -121b) have been reported as hosting inversions \citep{nymeyer11, sheppard17, arcangeli18, kreidberg18, evans17}.  Those three planets are highlighted on Figure~\ref{Tb_results}, and they tend to lie at the upper envelope with a high 4.5\,$\mu$m brightness temperature, albeit they are not decisively separated from the remainder of the sample.  However, the contribution functions of the 3.6- and 4.5\,$\mu$m bands are often overlapping (see Figure~12 of \citealp{kreidberg18}), so temperature inversions will tend to raise both the 3.6- and 4.5\,$\mu$m brightness temperatures.  In the case where the inversion extends over a broad range of pressure, planets will tend to move {\it along} the model track, rather than perpendicular to it.  The inverted Fortney model track illustrates this point: at high temperature it merges with the track for non-inverted models, but a given planet lies at a lower or higher position on the track depending on whether the temperature gradient is normal or inverted.  In order to move planets above and away from the model track (significantly brighter at 4.5\,$\mu$m), it is necessary to 'fine tune' the temperature inversion to affect the 4.5\,$\mu$m contribution function, while minimizing the impact on the 3.6\,$\mu$m contribution function.

We cannot exclude the possibility that multiple mechanisms are at play when accounting for our results.  One possibility is Burrows-like strong absorption (see Sec.~\ref{subsec_deviations}) for planets with equilibrium temperatures below $\sim 2000K$, coupled with blackbody-like behavior for the hottest planets due to the water dissociation and chemistry/opacity issues discussed by \citet{parmentier18b} and \citet{lothringer18}.  Another possibility is a metallicity effect that comes into play at low temperature as discussed by \citet{kammer15}, as well as possible temperature inversions for the hottest planets.  Also, emission in CO due to mass loss \citep{bell19} could increase Tb(4.5) for the most strongly irradiated planets.  However, we prefer the simplicity of a single hypothesis to account for the total effect that we observe.  As regards temperature inversions, we do not think they play a major role in our results, for several reasons:  1) Inversions have to be fine-tuned to raise planets relative to the model track, 2) the three nominally inverted planets on Figure~\ref{Tb_results} are not significantly separated from the rest of the sample, and 3) inversions are unlikely to be sufficiently prevalent to affect the brightness temperature ratio over the large range of temperature illustrated on Figure~\ref{slope_fit_ratio}.  

We point out that \textit{Spitzer's} Tb(4.5) measurement can be a significant factor driving retrievals toward an atmospheric temperature inversion (e.g., for WASP-18b, \citealp{nymeyer11, sheppard17}). Given a systematic tendency for hotter planets to be relatively brighter than the models at 4.5\,$\mu$m, together with random noise, some of the hottest planets may then reach a threshold where the retrieval codes react by requiring a temperature inversion for planets at the upper end of the distribution in Tb(4.5). Our 'big picture' data suggest that the primary difference between the models and the real planets is systematic over a large range of temperature, rather than inversions in some of the hottest planets. 

We suggest that Figure~\ref{slope_fit_ratio} requires a pervasive difference between the models and the real planets, systematically affecting the temperature versus optical depth structure as a function of equilibrium temperature.  One possibility is a difference in opacities between the planets and the models.  Another possibility is the effect of a vigorous zonal circulation on the radial temperature gradient (i.e., 3-D versus 1-D models) is one possibility.  In that respect, the greater efficiency of heat redistribution on cooler versus hotter planets (Figure~\ref{fig_heat}) is potentially an important factor.  Other possibilities include systematic changes in haze opacity (particle size, composition, and height) as a function of equilibrium temperature, and height gradients in the relative mixing ratios of CO and water vapor (chemical equilibrium, or not).  The physics underlying this systematic trend can hopefully be clarified using spectroscopy by JWST.

\section{Summary} \label{sec_summary}

In this paper we have investigated the emergent spectra of transiting hot Jupiters, using their secondary eclipses as observed in the two warm \textit{Spitzer} bands at 3.6- and 4.5\,$\mu$m.  We report eclipse depths for twenty seven previously unobserved planets, and we re-analyze eclipses of 9 previously observed planets in order to compare and relate our results to published work.  Our new planets include highly irradiated worlds such as KELT-7b, WASP-87b, WASP-76b, and WASP-64b, as well as others that are important targets for JWST, such as WASP-62b.  We also analyze \textit{Spitzer} {\it transits} of KELT-7, WASP-62, and WASP-74, in order to improve the precision of their orbital periods (Section~\ref{ssec_ephem}).  Our \textit{Spitzer} eclipse fits (Section~\ref{ssec_depths}) utilize photometry extracted using four different methods (Section~\ref{sec_obs}), each with multiple aperture sizes, and a pixel-level decorrelation method to correct instrumental effects and thereby select the optimum values of eclipse depth.  We investigate and discuss the statistical properties of our fitted eclipse depths (Section~\ref{ssec_checks}), including a comparison to the magnitude of the photon noise, analysis of the Allan deviation slope, and comparison to eclipse depths for the 9 planets previously published.  

The orbital phase of a secondary eclipse is sensitive to non-zero orbital eccentricities, and we investigate those phases for our sample of planets (Section~\ref{sec_phase}).  We find statistical evidence that eclipses tend to increasingly deviate from phase 0.5, the deviation increasing with orbital period in the range of our sample (periods 0.8 to 5.3 days), indicating an increasing lack of orbital circularization.  We conclusively find a slightly eccentric orbit for  WASP-62b ($e\cos{\omega} = 0.00614\pm{0.00064}$, Section~\ref{sec_phase}), that lies in the continuous viewing zone of JWST.  The eclipse of that planet occurs about 23 minutes later than orbital phase 0.5, and that delay is significant for planning of JWST observations.  Even for circular orbits, the phase of secondary eclipse is predicted to be offset from 0.5 due to temperature structure on the exoplanetary disk \citep{williams06}. Excluding planets with notably eccentric orbits, our sample has an average eclipse phase over both \textit{Spitzer} wavelengths that is centered on 0.5 to a precision of about $\pm23$\,seconds.  We do not detect a time offset because our precision is comparable to the offset predicted by \citet{williams06}, but we do exclude some of the larger values that they modeled. Our precision on the average eclipse phase of our sample is modestly poorer than the offset successfully measured for HD\,189733b by \citet{agol10}.  We project that a complete sample of \textit{Spitzer} eclipses (all planets observed), especially with improved precision in their orbital ephemerides, would be sufficient to detect the offset for the 'average planet', thereby extending the result from \citet{agol10} to the larger sample.   

We apply corrections for dilution of eclipse depths by stellar companions to some systems (Sec.~\ref{ssec_dilution}), and then convert the eclipse depths to brightness temperatures in each \textit{Spitzer} band (Section~\ref{sec_bright}), using ATLAS model atmospheres for the host stars \citep{kurucz}.  We use those brightness temperatures to investigate heat redistribution on the day sides of the planets (Section~\ref{sec_heat}), following the approach of \citet{cowan}.  We find that planets whose calculated maximum day size temperature exceeds $\sim 2200$K are well described by an observed brightness temperature consistent with zero albedo and redistribution of stellar irradiance uniformly over the day side.  About 35\% of planets whose calculated maximum temperature falls between $\sim\,1700$K and $\sim\,2200$K require non-zero albedos, even if their redistribution of stellar irradiance is uniform over the entire planet. Six planets in our sample have published \textit{Spitzer} phase curves, and these planets are typical of the entire sample, and consistent with uniform redistribution of stellar irradiance over the day side.  

To investigate the emergent day side spectra of our planets, we invoke a statistical approach whereby we compare brightness temperatures in the two \textit{Spitzer} bands, and seek trends for the entire sample (Section~\ref{subsec_stat}).  We compare the observed brightness temperatures (Tb) to two sets of well documented model atmospheres, from Adam Burrows and Jonathan Fortney (Section~\ref{subsec_models}), both based on cloudless atmospheres with solar abundances. Those models differ in the amplitude of their absorption features due to differences in their temperature structures, with the Burrows models predicting stronger absorptions than the Fortney models.  We also compare the observed brightness temperatures to blackbody planets (Section~\ref{subsec_deviations}), for which the day side brightness temperatures would be equal in the two \textit{Spitzer} bands.  In the Tb(4.5) versus Tb(3.6) plane, the observed planets seem to slope more steeply than a blackbody, with the hottest planets being brighter at 4.5 relative to 3.6, and the cooler planets being fainter at 4.5 relative to 3.6.  {While that tendency is not statistically secure, it did motivate us to investigate a similar trend, that we find to be robust (see below). 

Comparing the observed brightness temperatures to models, we find that the Burrows and Fortney models bracket 60\% of the observed planets in Tb(4.5), with the Fortney models lying near the center in Tb(4.5), and the Burrows models at the lower envelope. Because molecular absorptions are stronger in the 4.5\,$\mu$m band than at 3.6\,$\mu$m, that bracketing thereby constrains the average amplitude of absorption features in the day side spectra of our planets.  

Our most intriguing result is that the ratio of Tb(4.5) to Tb(3.6) increases with equilibrium temperature, and we show that this trend is statistically significant (Section~\ref{subsec_slope}), and is not due to selection effects (Section~\ref{subsec_select}).  Adding lower temperature planets (800 to 1200K) from \citet{kammer15} and \citet{wallack18}, we find that the ratio of Tb(4.5) to Tb(3.6) increases by 100$\pm24$ ppm for each 1K increase in equilibrium temperature from 800K to 2500K.  No existing model predicts this trend over such a large range of temperature. While it could in principle be due to a combination of effects such as temperature inversions in the hotter planets of the sample, coupled with stronger-than-modeled molecular absorption for the cooler planets, we advance the simple hypothesis (Section~\ref{subsec_structure}) that it represents a structural difference in the atmospheric temperature profile between the real planetary atmospheres compared to models.

\section{Acknowledgements}
We thank the staff of the Spitzer Space Telescope for their help in planning, and their careful scheduling and execution of the observations.   We also thank an anonymous referee and the statistical editor for comments that significantly improved this paper.  This work was supported by NASA ADAP grant NNX16AF34G.



\clearpage

\begin{figure}\centering
    \includegraphics[width=7in, height=7in]{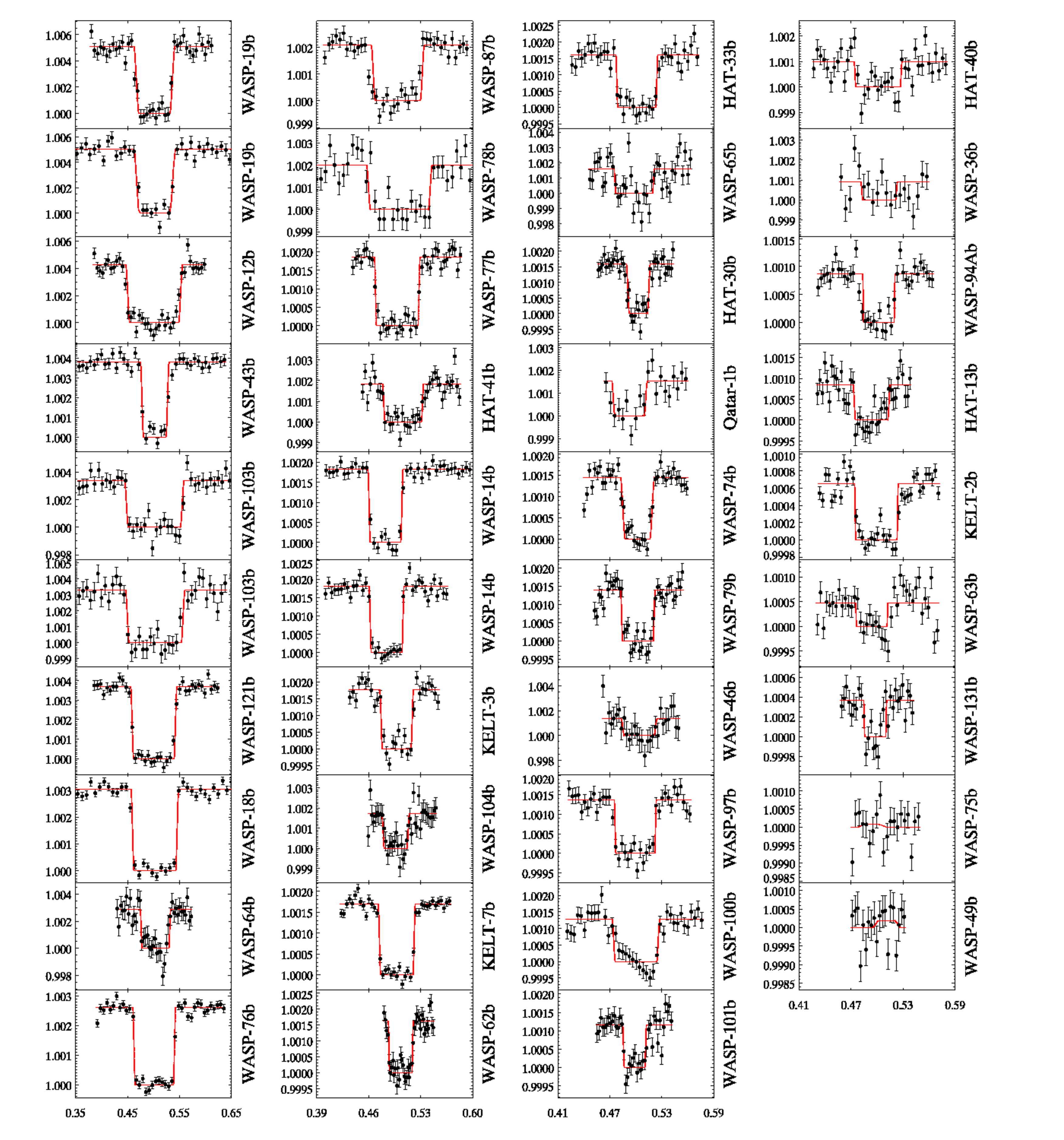}
    \begin{minipage}{6in}
    \caption{Eclipses at 3.6\,$\mu$m for all hot Jupiters analyzed in this paper. The abscissa for all plots is orbital phase and the ordinate is relative flux.  The eclipses are sorted by deepest to shallowest eclipse depth, going top to bottom and left to right. The data are binned for clarity, with between 20 and 40 points per dataset. The fitted eclipse is overplotted in red. The error bars are the scatter in each individual bin. The planet names are to the right of each plot. Note that the x-axis scale changes between columns and y-axis scale changes between each eclipse.  All eclipses are nominally detected (i.e., they have positive depths near the expected phase), except for WASP-75b and -49b (lowest right), where negative eclipse depths are derived. Considering the planets with positive eclipses, the ratio of eclipse depth to its random error varies from 1.6 (WASP-36) to 48 (WASP-18), and the median is 15.}
    \end{minipage}
\label{all36}    
\end{figure}
\clearpage

\begin{figure}\centering
    \includegraphics[width=7in, height=7in]{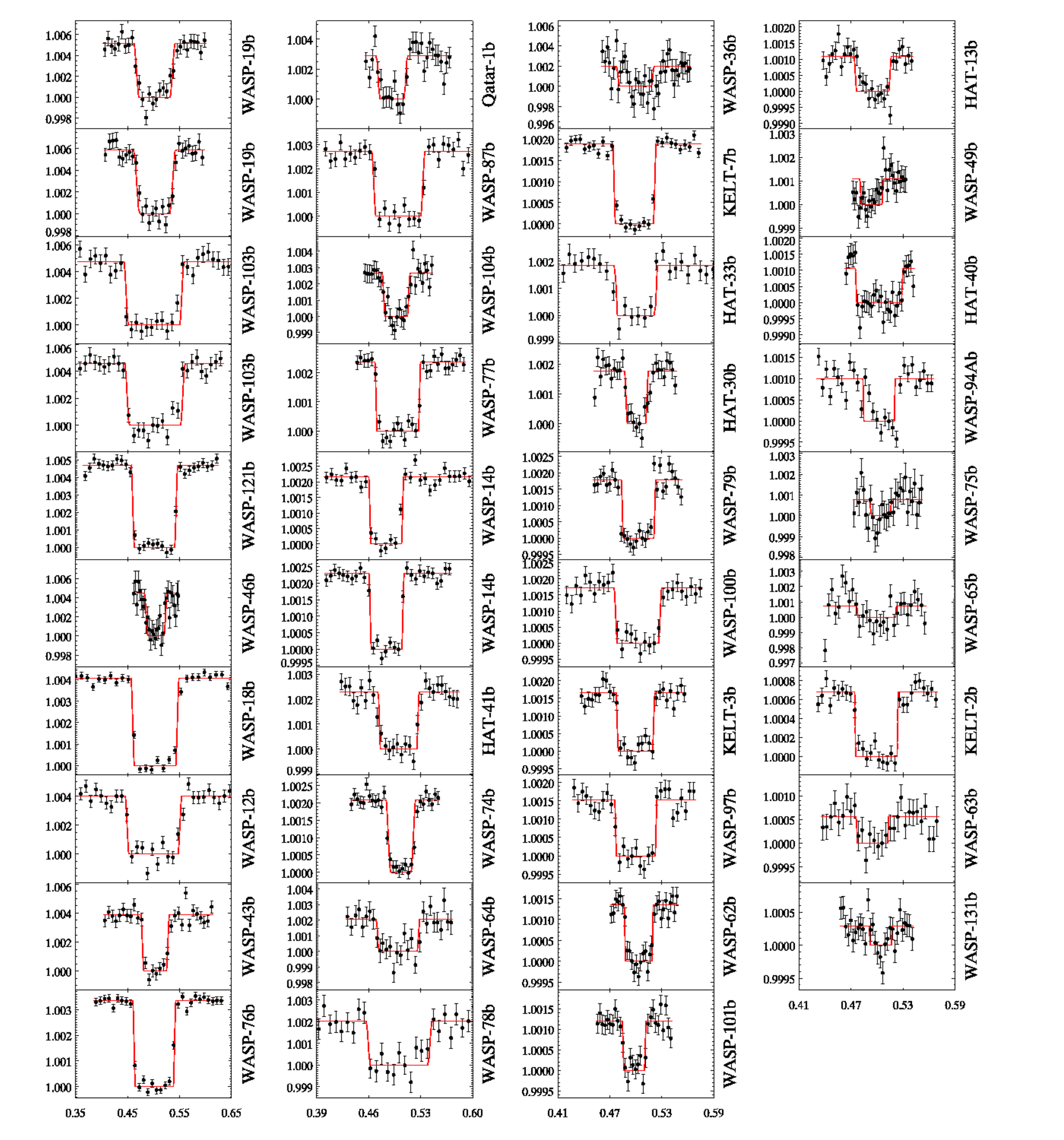}
    \begin{minipage}{6in}
    \caption{Eclipses at 4.5\,$\mu$m for all hot Jupiters analyzed in this paper, similar to the 3.6\,$\mu$m eclipses shown in Figure~\ref{all36}.  All eclipses are nominally detected, and the ratio of eclipse depth to its random error varies from 1.1 (WASP-75) to 41 (WASP-18), and the median is 12.}
    \end{minipage}
\label{all45}    
\end{figure}

\begin{table*} 
\centering
\advance\leftskip-2.1cm
\caption{Eclipse depths (ED) in contrast units of parts-per-million, normalized to the flux from the host star.  These are 'as observed', without dilution corrections applied. Dilution correction factors are given in Table~\ref{dilute_table}.  The type of fit to the photometry is encoded as: temporal baseline (L=linear, Q=quadratic, E=exponential), centroiding method (C=center of light, G=2-D Gaussian fit), photometric aperture type (F=fixed radii, V= variable, using the noise-pixel formulation), and the number of minutes trimmed from the start of the observations.  For example LGF30 means a linear baseline, Gaussian centroiding, fixed radius aperture, and 30 minutes trimmed from the start of the observations. The 'bin' column lists the bin size used in the PLD solutions (see text). 'Ratio' is the ratio of scatter in the unbinned residuals, divided by the photon noise. 'Slope' is the slope of the relation between the log of standard deviation of the residuals using multiple bin sizes, versus the square root of the bin size.  The AOR is the Astronomical Observation Request number that uniquely identifies the data we used from the Spitzer Heritage Archive.}
\resizebox{\textwidth}{!}{
\begin{tabular}{lllllllllllll}
Planet Name & AOR & \multicolumn{1}{c}{3.6 ED (ppm)} & \multicolumn{1}{c}{Fit Type} & \multicolumn{1}{c}{Bin} & \multicolumn{1}{c}{Ratio} & 
\multicolumn{1}{c}{Slope}  & \multicolumn{1}{c}{AOR}  & \multicolumn{1}{c}{4.5 ED (ppm)} &  Fit Type & Bin & Ratio & Slope \\
\hline
            HAT-13 &   38808320 &  851$\pm$107 &  ECV30 &  120 &  1.28 & -0.42 &   38808832 & 1090$\pm$124 &  LCF30 &   96 &  1.17 & -0.50 \\
            HAT-30 &   42612736 & 1584$\pm$107 &  LCV60 &   28 &  1.11 & -0.45 &   42613504 & 1762$\pm$147 &  LCF60 &   16 &  1.21 & -0.45 \\
            HAT-33 &   62151424 & 1603$\pm$127 &  QGF45 &   18 &  1.15 & -0.47 &   51838720 & 1835$\pm$199 &  LCF45 & 1344 &  1.25 & -0.48 \\
            HAT-40 &   51832064 &  988$\pm$168 &  LGF30 &  544 &  1.30 & -0.40 &   62151936 & 1057$\pm$145 &  LGF30 &    2 &  1.08 & -0.51 \\
            HAT-41 &   51840512 & 1829$\pm$319 &  QGF30 &  512 &  1.23 & -0.39 &   51838464 & 2278$\pm$177 &  LGF30 &  640 &  1.20 & -0.45 \\
            KELT-2 &   51835136 &  650$\pm$ 38 &  QGV30 &   34 &  1.11 & -0.40 &   51833600 &  678$\pm$ 47 &  QGV30 &   26 &  1.09 & -0.49 \\
            KELT-3 &   51815936 & 1766$\pm$ 97 &  LGF30 &  480 &  1.28 & -0.40 &   51842048 & 1656$\pm$104 &  LGF30 &  320 &  1.16 & -0.45 \\
            KELT-7 &   62154496 & 1688$\pm$ 46 &  QGV30 &   34 &  1.05 & -0.43 &   62155520 & 1896$\pm$ 57 &  QGF30 &   46 &  1.12 & -0.46 \\
           Qatar-1 &   51819776 & 1511$\pm$455 &  LGF45 &  336 &  1.36 & -0.49 &   51816960 & 2907$\pm$415 &   LGF0 &  544 &  1.58 & -0.49 \\
           WASP-12 &   48014848 & 4247$\pm$243 &  QGF30 &   28 &  1.15 & -0.40 &   48015872 & 3996$\pm$171 &  LCF30 &   10 &  1.21 & -0.46 \\
           WASP-14 &   45426944 & 1816$\pm$ 67 &  QCV30 &   60 &  1.10 & -0.46 &   45426688 & 2161$\pm$ 88 &  QCF30 &   12 &  1.19 & -0.43 \\
           WASP-14 &   45427968 & 1798$\pm$ 59 &  LCV30 &    8 &  1.11 & -0.49 &   45428992 & 2284$\pm$ 90 &  QCF30 &   54 &  1.19 & -0.40 \\
           WASP-18 &   38805760 & 3037$\pm$ 62 &  QCF30 &   24 &  1.10 & -0.46 &   40269312 & 4033$\pm$ 97 &  QCF30 &   26 &  1.10 & -0.49 \\
           WASP-19 &   43970048 & 5016$\pm$259 &  QCV30 &  120 &  1.31 & -0.49 &   43970560 & 5081$\pm$392 &  QGF30 &    6 &  1.17 & -0.48 \\
           WASP-19 &   43970048 & 5070$\pm$233 &  QCF30 &    4 &  1.14 & -0.53 &   43970560 & 5848$\pm$544 &  QGF30 &  108 &  1.17 & -0.48 \\
           WASP-36 &   51829504 &  913$\pm$578 &  LGF30 &   18 &  1.25 & -0.47 &   51827456 & 1948$\pm$544 &  LGF30 &   62 &  1.16 & -0.51 \\
           WASP-43 &   42614272 & 3773$\pm$138 &  QGV30 &   32 &  1.22 & -0.45 &   42615040 & 3866$\pm$195 &  LCF30 &   14 &  1.23 & -0.45 \\
           WASP-46 &   51823872 & 1360$\pm$701 &  LGF30 &   50 &  1.38 & -0.49 &   51821568 & 4446$\pm$589 &  LGV30 &   92 &  1.34 & -0.54 \\
           WASP-49 &   51828480 & -189$\pm$265 &  LCF60 &  184 &  1.21 & -0.40 &   51826688 & 1073$\pm$336 &  LCF60 &  304 &  1.22 & -0.51 \\
           WASP-62 &   51823360 & 1616$\pm$146 &  QGF45 &  352 &  1.16 & -0.46 &   51821056 & 1359$\pm$130 &  QGF45 &  448 &  1.15 & -0.47 \\
           WASP-63 &   51835904 &  486$\pm$ 96 &  LGF30 &   42 &  1.23 & -0.43 &   51834112 &  560$\pm$130 &  LGF30 & 1024 &  1.18 & -0.50 \\
           WASP-64 &   51816704 & 2859$\pm$270 &  LGF30 &   18 &  1.24 & -0.44 &   51842560 & 2071$\pm$471 &  LGF30 &  108 &  1.16 & -0.48 \\
           WASP-65 &   51828224 & 1587$\pm$245 &  LGV30 &    2 &  1.67 & -0.40 &   51826432 &  724$\pm$318 &  LGV30 &    2 &  1.56 & -0.46 \\
           WASP-74 &   62170880 & 1446$\pm$ 66 &  LGV30 &   34 &  1.07 & -0.44 &   62171904 & 2075$\pm$100 &  LGF30 &  136 &  1.10 & -0.50 \\
           WASP-75 &   51826176 &  -86$\pm$290 &    QCV &    2 &  1.44 & -0.44 &   51824384 &  452$\pm$399 &   LCF0 &  168 &  1.29 & -0.49 \\
           WASP-76 &   58239232 & 2597$\pm$ 63 &  QGF30 &   18 &  1.11 & -0.43 &   58238720 & 3344$\pm$ 82 &  QCF30 &   96 &  1.11 & -0.48 \\
           WASP-77 &   51820544 & 1845$\pm$ 94 &  QGF30 &    4 &  1.19 & -0.46 &   51818496 & 2362$\pm$127 &  QGF30 &   40 &  1.20 & -0.47 \\
           WASP-78 &   51833088 & 2001$\pm$218 &  LGV30 &   80 &  1.54 & -0.45 &   51830528 & 2013$\pm$351 &  QGF30 &  168 &  1.14 & -0.49 \\
           WASP-79 &   51841536 & 1394$\pm$ 88 &  LCF60 &   64 &  1.19 & -0.41 &   51839488 & 1783$\pm$106 &  LCF60 &  288 &  1.15 & -0.48 \\
           WASP-87 &   62173952 & 2077$\pm$127 &  LCV45 &   88 &  1.10 & -0.45 &   62174464 & 2705$\pm$137 &  LGF45 &    8 &  1.21 & -0.49 \\
          WASP-94A &   62174976 &  867$\pm$ 59 &  LCV30 &   10 &  1.16 & -0.42 &   62176000 &  995$\pm$ 93 &  LGF30 &   12 &  1.23 & -0.42 \\
           WASP-97 &   62177024 & 1359$\pm$ 84 &  QCV30 &    4 &  1.09 & -0.48 &   62177536 & 1534$\pm$101 &  LCF30 &   12 &  1.12 & -0.46 \\
          WASP-100 &   62156544 & 1267$\pm$ 98 &  LGF30 &   30 &  1.19 & -0.40 &   62157056 & 1720$\pm$119 &  LGF30 &   18 &  1.13 & -0.46 \\
          WASP-101 &   62157568 & 1161$\pm$111 &  LGV30 &    8 &  1.26 & -0.43 &   62158592 & 1194$\pm$113 &  LGV30 &   10 &  1.13 & -0.47 \\
          WASP-103 &   53518080 & 3294$\pm$193 &  LGV30 &    4 &  1.13 & -0.50 &   53513472 & 4552$\pm$369 &  QCF30 &   14 &  1.14 & -0.47 \\
          WASP-103 &   53519104 & 3350$\pm$218 &  QCV30 &    4 &  1.26 & -0.51 &   53514240 & 4711$\pm$339 &  QCF30 &    8 &  1.12 & -0.47 \\
          WASP-104 &   62179584 & 1709$\pm$195 &  LGF30 &   18 &  1.27 & -0.45 &   62180864 & 2643$\pm$303 &  QGF30 &   12 &  1.18 & -0.47 \\
          WASP-121 &   62159616 & 3685$\pm$114 &  QCV30 &   26 &  1.08 & -0.50 &   62160640 & 4684$\pm$121 &  LCF30 &   12 &  1.11 & -0.48 \\
          WASP-131 &   62162688 &  364$\pm$ 96 &  QCV30 &   68 &  1.16 & -0.45 &   62163712 &  289$\pm$ 80 &  LCF30 &    8 &  1.16 & -0.42 \\

\end{tabular}
}
\label{eclipse_depths}
\end{table*}

\begin{table*}
\centering
\caption{Transit times in BJD(TDB), transit depths, and updated orbital periods for KELT-7b, WASP-62b, and WASP-74b, based on the transits discussed in Section~\ref{ssec_ephem}.  The values of $T_0$ for all three planets are repeated from \citet{bieryla15}, \citet{hellier12}, and \citet{hellier15}, but converted to TDB as needed. }
\begin{tabular}{lllll}
Planet &  3.6\,$\mu$m time & 4.5\,$\mu$ time  & $T_0$ BJD(TDB) & Period (days) \\
\hline
KELT-7b & $2457749.95953\pm{0.00016}$ & $2457758.16446\pm{0.00019}$ & $2456355.229809\pm{0.000198}$ & $2.73476468\pm{0.00000046}$ \\
WASP-62b & $2457717.23121\pm{0.00021}$ & $2457730.46660\pm{0.00024}$ & $2455855.39272\pm{0.00027}$  & $4.41193897\pm{0.00000074}$ \\
WASP-74b  & $2457768.16637\pm{0.00024}$ & $2457770.30472\pm{0.00029}$ & $2456506.8926\pm{0.0002}$  & $2.13775257\pm{0.00000046}$ \\
\end{tabular}
\label{ephem_table}
\end{table*}

\begin{table*}
\centering
\caption{\textit{Spitzer} transit depths ($R_p^2/R_s^2$, in ppm) for KELT-7b, WASP-62b, and WASP-74b.}
\begin{tabular}{lll}
Planet &  3.6\,$\mu$m &  4.5\,$\mu$m   \\
\hline
KELT-7b & 7925$\pm$62 &  8092$\pm$36 \\
WASP-62b &  12189$\pm$101  &  12250$\pm$87 \\
WASP-74b  &  9044$\pm$56  &  9197$\pm$43  \\
\end{tabular}
\label{tdepth_table}
\end{table*}

\begin{table*}
\centering\caption{Dilution corrections for secondary eclipse depth at both \textit{Spitzer} wavelengths.  The "as measured" eclipse depths listed in Table~\ref{eclipse_depths} were multiplied by these factors before they were used in the analyses reported in Sec.~7, and for the brightness temperatures listed in Table~\ref{temp_table}.}
\begin{tabular}{lll}
Planet & 3.6\,$\mu$m factor & 4.5\,$\mu$m factor \\
\hline
HAT-P-30b &  1.0121  & 1.0117 \\
HAT-P-33b &  1.0377  & 1.0332 \\
HAT-P-41b &  1.0069  & 1.0111 \\
KELT-2b  &   1.137   & 1.123  \\ 
KELT-3b   &  1.0125  & 1.0127  \\
WASP-12b  &  1.1101  & 1.0983 \\
WASP-36b  &  1.0015  & 1.0026 \\
WASP-49b  &  1.0130  & 1.0124 \\
WASP-76b  &  1.1470  & 1.1250 \\ 
WASP-77b  &  1.0929  & 1.0530 \\
WASP-87b  &  1.0014  & 1.0011 \\
WASP-103b &  1.1700  & 1.1490 \\
\end{tabular}
\label{dilute_table}
\end{table*}

\begin{table*}
\centering
\caption{Central phases and times of the secondary eclipses.  The ephemeris source column gives the reference used to calculate the orbital phase from the BJD(TDB) times.  The phases are 'as observed' and have not been corrected for light travel time across the orbit.  The errors in eclipse phase are purely due to the eclipse observations and do not include imprecision in the orbital ephemeris.  Note that our analysis in Sec.~\ref{sec_phase} {\it does} include uncertainty in the orbital ephemeris when analyzing the properties of the eclipse phases.}
\resizebox{\textwidth}{!}{%
\begin{tabular}{lllllll}
Planet  &  3.6\,$\mu$m Phase &  3.6\,$\mu$m BJD(TDB)  & 4.5\,$\mu$m Phase & 4.5\,$\mu$m BJD(TDB)  & Ephemeris source \\
\hline
       HAT-13 & 0.49378 $\pm$ 0.00120 &  55326.70691 $\pm$ 0.00351 & 0.49495 $\pm$ 0.00110 &  55355.87271 $\pm$ 0.00319 &              Southworth+2012 \\ 
      HAT-30 & 0.50284 $\pm$ 0.00073 &  55930.06169 $\pm$ 0.00205 & 0.50068 $\pm$ 0.00086 &  55944.10866 $\pm$ 0.00243 &             Maciejewski+2016 \\ 
      HAT-33 & 0.50109 $\pm$ 0.00070 &  57784.53823 $\pm$ 0.00243 & 0.50025 $\pm$ 0.00144 &  57027.10001 $\pm$ 0.00501 &                 Hartman+2011 \\ 
      HAT-40 & 0.49829 $\pm$ 0.00096 &  57058.96841 $\pm$ 0.00428 & 0.49815 $\pm$ 0.00070 &  57705.26801 $\pm$ 0.00312 &                 Hartman+2012 \\ 
      HAT-41 & 0.50689 $\pm$ 0.00098 &  57008.45732 $\pm$ 0.00264 & 0.50074 $\pm$ 0.00121 &  57021.91098 $\pm$ 0.00325 &                 Hartman+2012 \\ 
      KELT-2 & 0.49946 $\pm$ 0.00040 &  57009.21971 $\pm$ 0.00163 & 0.49952 $\pm$ 0.00041 &  57017.44755 $\pm$ 0.00170 &                  Beatty+2012 \\ 
      KELT-3 & 0.50691 $\pm$ 0.00059 &  57060.22792 $\pm$ 0.00158 & 0.50822 $\pm$ 0.00063 &  57057.52805 $\pm$ 0.00171 &                  Pepper+2013 \\ 
      KELT-7 & 0.50019 $\pm$ 0.00024 &  57737.65388 $\pm$ 0.00067 & 0.50022 $\pm$ 0.00026 &  57754.06256 $\pm$ 0.00071 &           Bieryla+2015;Table \\ 
     Qatar-1 & 0.49900 $\pm$ 0.00499 &  56987.42464 $\pm$ 0.00709 & 0.49806 $\pm$ 0.00186 &  56993.10340 $\pm$ 0.00264 &                 Collins+2017 \\ 
     WASP-12 & 0.49923 $\pm$ 0.00101 &  56638.88641 $\pm$ 0.00110 & 0.49784 $\pm$ 0.00129 &  56642.15916 $\pm$ 0.00141 &                    Chan+2011 \\ 
     WASP-14 & 0.48310 $\pm$ 0.00043 &  56033.05283 $\pm$ 0.00096 & 0.48410 $\pm$ 0.00043 &  56042.03013 $\pm$ 0.00096 &                    Wong+2014 \\ 
     WASP-14 & 0.48461 $\pm$ 0.00035 &  56035.30000 $\pm$ 0.00078 & 0.48454 $\pm$ 0.00042 &  56044.27490 $\pm$ 0.00093 &                    Wong+2014 \\ 
     WASP-18 & 0.50045 $\pm$ 0.00038 &  55220.83391 $\pm$ 0.00035 & 0.50083 $\pm$ 0.00040 &  55432.66092 $\pm$ 0.00037 &              Southworth+2009 \\ 
     WASP-19 & 0.50010 $\pm$ 0.00105 &  55776.76928 $\pm$ 0.00083 & 0.49982 $\pm$ 0.00152 &  55787.02396 $\pm$ 0.00120 &                    Wong+2016 \\ 
     WASP-19 & 0.49962 $\pm$ 0.00092 &  55777.55774 $\pm$ 0.00073 & 0.50011 $\pm$ 0.00160 &  55787.81303 $\pm$ 0.00126 &                    Wong+2016 \\ 
     WASP-36 & 0.50140 $\pm$ 0.00412 &  57055.70407 $\pm$ 0.00634 & 0.49832 $\pm$ 0.00368 &  57063.38618 $\pm$ 0.00566 &                 Mancini+2015 \\ 
     WASP-43 & 0.50033 $\pm$ 0.00070 &  55773.31778 $\pm$ 0.00057 & 0.50101 $\pm$ 0.00100 &  55772.50487 $\pm$ 0.00082 &               Stevenson+2017 \\ 
     WASP-46 & 0.50434 $\pm$ 0.00161 &  57000.77359 $\pm$ 0.00230 & 0.50298 $\pm$ 0.00161 &  57005.06275 $\pm$ 0.00230 &                Anderson+2012 \\ 
     WASP-49 & 0.50695 $\pm$ 0.00181 &  57003.47383 $\pm$ 0.00502 & 0.49379 $\pm$ 0.00131 &  57011.78245 $\pm$ 0.00364 &                   Lendl+2012 \\ 
     WASP-62 & 0.50421 $\pm$ 0.00052 &  56991.48560 $\pm$ 0.00230 & 0.50390 $\pm$ 0.00053 &  57062.07524 $\pm$ 0.00232 &      Hellier+2012;Brown+2017 \\ 
     WASP-63 & 0.49745 $\pm$ 0.00221 &  57013.97577 $\pm$ 0.00966 & 0.49456 $\pm$ 0.00144 &  57035.85357 $\pm$ 0.00630 &                 Hellier+2012 \\ 
     WASP-64 & 0.50208 $\pm$ 0.00135 &  57019.80703 $\pm$ 0.00213 & 0.50035 $\pm$ 0.00186 &  57015.08443 $\pm$ 0.00292 &                  Gillon+2013 \\ 
     WASP-65 & 0.49831 $\pm$ 0.00114 &  57047.96638 $\pm$ 0.00263 & 0.49977 $\pm$ 0.00493 &  57050.28117 $\pm$ 0.01139 &       Gomez-Maqueo-Chew+2013 \\ 
     WASP-74 & 0.50029 $\pm$ 0.00057 &  57769.23614 $\pm$ 0.00123 & 0.50217 $\pm$ 0.00051 &  57797.03093 $\pm$ 0.00109 &           Hellier+2015;Table \\ 
     WASP-75 & 0.49435 $\pm$ 0.00725 &  57055.88764 $\pm$ 0.01801 & 0.49626 $\pm$ 0.00444 &  57058.37657 $\pm$ 0.01103 &       Gomez-Maqueo-Chew+2013 \\ 
     WASP-76 & 0.49935 $\pm$ 0.00032 &  57469.79003 $\pm$ 0.00058 & 0.49945 $\pm$ 0.00034 &  57480.64953 $\pm$ 0.00062 &                    West+2016 \\ 
     WASP-77 & 0.49892 $\pm$ 0.00052 &  56975.47418 $\pm$ 0.00070 & 0.49959 $\pm$ 0.00056 &  56978.19514 $\pm$ 0.00076 &                  Maxted+2013 \\ 
     WASP-78 & 0.50180 $\pm$ 0.00253 &  56986.26549 $\pm$ 0.00551 & 0.50142 $\pm$ 0.00203 &  57005.84125 $\pm$ 0.00442 &                 Smalley+2012 \\ 
     WASP-79 & 0.50057 $\pm$ 0.00071 &  56993.71597 $\pm$ 0.00259 & 0.50133 $\pm$ 0.00062 &  57004.70593 $\pm$ 0.00227 &      Smalley+2012;Brown+2017 \\ 
     WASP-87 & 0.49965 $\pm$ 0.00090 &  57690.40719 $\pm$ 0.00152 & 0.50037 $\pm$ 0.00092 &  57692.09119 $\pm$ 0.00156 &                Anderson+2014 \\ 
     WASP-94A & 0.50213 $\pm$ 0.00096 &  57773.30115 $\pm$ 0.00378 & 0.50231 $\pm$ 0.00106 &  57777.25201 $\pm$ 0.00420 &          Neveu-VanMalle+2014 \\ 
     WASP-97 & 0.49935 $\pm$ 0.00054 &  57695.31527 $\pm$ 0.00113 & 0.49993 $\pm$ 0.00067 &  57699.46200 $\pm$ 0.00138 &                 Hellier+2014 \\ 
    WASP-100 & 0.50011 $\pm$ 0.00089 &  57698.45281 $\pm$ 0.00254 & 0.50235 $\pm$ 0.00086 &  57704.15794 $\pm$ 0.00245 &                 Hellier+2014 \\ 
    WASP-101 & 0.49837 $\pm$ 0.00066 &  57762.12752 $\pm$ 0.00236 & 0.49792 $\pm$ 0.00075 &  57780.05450 $\pm$ 0.00269 &                 Hellier+2014 \\ 
    WASP-103 & 0.50119 $\pm$ 0.00148 &  57171.80783 $\pm$ 0.00137 & 0.50062 $\pm$ 0.00183 &  57163.47739 $\pm$ 0.00169 &              Southworth+2015 \\ 
    WASP-103 & 0.49947 $\pm$ 0.00140 &  57170.88069 $\pm$ 0.00129 & 0.49882 $\pm$ 0.00144 &  57162.55018 $\pm$ 0.00133 &              Southworth+2015 \\ 
    WASP-104 & 0.49673 $\pm$ 0.00124 &  57851.68947 $\pm$ 0.00218 & 0.49749 $\pm$ 0.00100 &  57856.95704 $\pm$ 0.00176 &                   Smith+2014 \\ 
    WASP-121 & 0.49905 $\pm$ 0.00053 &  57783.77754 $\pm$ 0.00067 & 0.50034 $\pm$ 0.00055 &  57906.17204 $\pm$ 0.00070 &                  Delrez+2016 \\ 
    WASP-131 & 0.49787 $\pm$ 0.00126 &  57917.69235 $\pm$ 0.00670 & 0.50425 $\pm$ 0.00398 &  57912.40429 $\pm$ 0.02118 &                 Hellier+2017 \\ 

\end{tabular}
}
\label{phase_table}
\end{table*}

\begin{table*}
\centering
\caption{Equilibrium temperatures and brightness temperatures in the \textit{Spitzer} bands, calculated as described in Sec.~\ref{sec_bright}.  Note that the eclipse depths listed in Table~\ref{eclipse_depths} were corrected for dilution (Table~\ref{dilute_table}) in the process of calculating these brightness temperatures.}
\begin{tabular}{llll}
Planet  &  Equilibrium temperature &  3.6\,$\mu$m Tb  & 4.5\,$\mu$m Tb  \\
\hline
               HAT-13 & 1653$\pm$  50 & 1810$\pm$ 229 & 1754$\pm$ 200 \\ 
              HAT-30 & 1718$\pm$  34 & 2087$\pm$ 140 & 1938$\pm$ 160 \\ 
              HAT-33 & 1855$\pm$ 148 & 2112$\pm$ 162 & 1990$\pm$ 209 \\ 
              HAT-40 & 1771$\pm$  38 & 2074$\pm$ 354 & 1887$\pm$ 259 \\ 
              HAT-41 & 1685$\pm$  58 & 1694$\pm$ 294 & 1622$\pm$ 125 \\ 
              KELT-2 & 1721$\pm$  36 & 1994$\pm$ 104 & 1782$\pm$ 111 \\ 
              KELT-3 & 1829$\pm$  42 & 2445$\pm$ 133 & 2132$\pm$ 133 \\ 
              KELT-7 & 2056$\pm$  31 & 2512$\pm$  69 & 2415$\pm$  73 \\ 
             Qatar-1 & 1422$\pm$  36 & 1410$\pm$ 425 & 1532$\pm$ 219 \\ 
             WASP-12 & 2546$\pm$  82 & 3329$\pm$ 172 & 2934$\pm$ 114 \\ 
             WASP-14 & 1893$\pm$  60 & 2302$\pm$  85 & 2256$\pm$  92 \\ 
             WASP-14 & 1893$\pm$  60 & 2292$\pm$  76 & 2319$\pm$  92 \\ 
             WASP-18 & 2416$\pm$  58 & 3057$\pm$  63 & 3323$\pm$  80 \\ 
             WASP-19 & 2099$\pm$  39 & 2451$\pm$ 127 & 2191$\pm$ 169 \\ 
             WASP-19 & 2099$\pm$  39 & 2465$\pm$ 114 & 2353$\pm$ 219 \\ 
             WASP-36 & 1705$\pm$  44 & 1336$\pm$ 844 & 1506$\pm$ 420 \\ 
             WASP-43 & 1444$\pm$  40 & 1781$\pm$  65 & 1537$\pm$  78 \\ 
             WASP-46 & 1663$\pm$  54 & 1435$\pm$ 740 & 2014$\pm$ 267 \\ 
             WASP-49 & 1320$\pm$  88 &  ---          & 1256$\pm$ 389 \\ 
             WASP-62 & 1432$\pm$  33 & 1955$\pm$ 177 & 1593$\pm$ 153 \\ 
             WASP-63 & 1536$\pm$  37 & 1547$\pm$ 308 & 1395$\pm$ 324 \\ 
             WASP-64 & 1674$\pm$ 169 & 2135$\pm$ 202 & 1607$\pm$ 366 \\ 
             WASP-65 & 1490$\pm$  45 & 1833$\pm$ 284 & 1179$\pm$ 518 \\ 
             WASP-74 & 1922$\pm$  46 & 2049$\pm$  94 & 2161$\pm$ 105 \\ 
             WASP-75 & 1710$\pm$  39 &  ---          & 1112$\pm$ 983 \\ 
             WASP-76 & 2190$\pm$  43 & 2669$\pm$  57 & 2747$\pm$  60 \\ 
             WASP-77 & 1677$\pm$  28 & 1786$\pm$  84 & 1696$\pm$  87 \\ 
             WASP-78 & 2200$\pm$  41 & 3034$\pm$ 331 & 2763$\pm$ 483 \\ 
             WASP-79 & 1760$\pm$  51 & 1959$\pm$ 125 & 1948$\pm$ 117 \\ 
             WASP-87 & 2320$\pm$  62 & 2802$\pm$ 172 & 2988$\pm$ 152 \\ 
            WASP-94A & 1508$\pm$  75 & 1385$\pm$  95 & 1249$\pm$ 118 \\ 
             WASP-97 & 1545$\pm$  40 & 1772$\pm$ 111 & 1615$\pm$ 107 \\ 
            WASP-100 & 2207$\pm$ 170 & 2306$\pm$ 180 & 2429$\pm$ 168 \\ 
            WASP-101 & 1559$\pm$  38 & 1723$\pm$ 166 & 1524$\pm$ 145 \\ 
            WASP-103 & 2513$\pm$  49 & 2993$\pm$ 150 & 3268$\pm$ 231 \\ 
            WASP-103 & 2513$\pm$  49 & 2771$\pm$ 181 & 3066$\pm$ 221 \\ 
            WASP-104 & 1501$\pm$ 189 & 1716$\pm$ 197 & 1783$\pm$ 205 \\ 
            WASP-121 & 2366$\pm$  57 & 2490$\pm$  77 & 2562$\pm$  66 \\ 
            WASP-131 & 1463$\pm$  32 & 1397$\pm$ 369 & 1114$\pm$ 309 \\ 
\end{tabular}
\label{temp_table}
\end{table*}

\clearpage

\section{APPENDIX:  Notes for Some Individual Planets} \label{sec_notes}

{\bf HAT-P-13b} has been previously analyzed by \citet{buhler16} and \citet{hardy17}.  Like those investigations, we concur that the eclipse occurs slightly before phase 0.5, and thus the orbit is slightly eccentric.  Our phases agree especially well with \citet{buhler16}, but are also in reasonable agreement with \citet{hardy17}.  The previous investigations found somewhat discordant eclipse depths at 4.5\,$\mu$m: \citet{hardy17} derived $810\pm80$\,ppm, whereas \citet{buhler16} derived $1426\pm130$\,ppm.  Our value ($1090\pm124$\,ppm) is intermediate between them. 

{\bf HAT-P-30b} was announced by \citet{johnson11}, and the orbital parameters were updated by \citet{maciejewski16}.  Since the latter are more recent, we initially used those orbital parameters to generate the shape of the secondary eclipse curve that we fit to our \textit{Spitzer} data.  However, we found that the eclipse shape using the original orbital parameters (i.e., inclination, $a/R_s$, etc) from \citet{johnson11} gave much better agreement with our \textit{Spitzer} data.  We retained the orbital period and transit epoch as updated by \citet{maciejewski16}. Our dilution correction is based on our scattering fractions from the {\it Spitzer} photometry (see text, Section~\ref{ssec_dilution}), supplemented by a magnitude difference from \citet{evans18}.

{\bf HAT-P-33b} has a close companion star, entirely contained within {\it Spitzer's} point spread function.  Our dilution correction is based on the magnitude differences and temperatures from \citet{ngo15}.

{\bf KELT-2b} has a close companion, entirely contained within {\it Spitzer's} point spread function.  To calculate our dilution correction, we used data from \citet{beatty12}. 

{\bf Qatar-1b} has minimal eclipse baseline at 3.6\,$\mu$m before ingress due to the presence of a strong ramp and required trimming 45 minutes of initial data.  However, we found no significant ramp at 4.5\,$\mu$m, allowing us to use the full data without trimming.  The eclipse depths and phases reported here are slight updates from the values we previously published in \citet{garhart18}, but the differences are within the errors, and not significant for the emergent spectrum or the orbital dynamics.

{\bf WASP-12b} was analyzed by one of us (D.D.) for the eclipse timing results reported in \citet{patra}.  The updated eclipse times we list here agree with \citet{patra} to $<1\sigma$.  Note also that these eclipse data were observed \textit{Spitzer} program 90186 (P.I. = Kamen Todorov), and the eclipse depths are reported here for the first time.  In calculating the dilution correction, we used data from \citet{hebb09}, and \citet{bechter14} (also see \citealp{crossfield12}). 

{\bf WASP-46b} was observed in our Cycle-10 program that was Priority=3 for \textit{Spitzer}.  We accordingly used a minimum total duration in order to maximize the probability that the observations would be scheduled.  Together with a slightly late eclipse phase (possibly due to ephemeris error), the observed eclipse has minimal eclipse baseline at 3.6\,$\mu$m after egress.

{\bf WASP-49b} has a minimal eclipse baseline at 4.5\,$\mu$m before ingress due to the presence of a strong ramp, that required trimming 60 minutes of initial data.  \citet{lendl16} note the presence of a companion star at 2.2 arc-sec, and \citet{evans18} derived the temperature of the companion star, an M-dwarf.  We based our dilution correction on the 2MASS K-magnitudes for the primary star and companion, together with the companion temperature (3230K) from \citet{evans18}.

{\bf WASP-62b} was observed in our Cycle-10 program that was Priority=3 for \textit{Spitzer}.  We accordingly used a minimum total duration in order to maximize the probability that the observations would be scheduled.  Moreover, a relatively strong ramp at 3.6\,$\mu$m required trimming 45 minutes of data at 3.6\,$\mu$m. Nevertheless, good agreement in the phase of the eclipse in both bands reinforces our confidence in the eclipse depths as well as the phases.

{\bf WASP-74b} was announced by \citet{hellier15}, who derived an optical transit depth ($R_p^2/R_s^2$) of $9610\pm140$\,ppm, about 5\% larger than the Spitzer transit depths we give in Table~\ref{tdepth_table}. We suggest that much of the difference is due to the stellar limb darkening, since this transit is nearly grazing (impact parameter = 0.86, \citealp{hellier15}).  We used quadratic limb darkening at both Spitzer wavelengths, from \citet{claret13}, and the (linear, quadratic) coefficients we used are (0.0946, 0.1141) at 3.6\,$\mu$m, and (0.0798, 0.0963) at 4.5\,$\mu$m. 

{\bf WASP-75b} was observed in our Cycle-10 program that was Priority=3 for \textit{Spitzer}.  We accordingly used a minimum total duration in order to maximize the probability that the observations would be scheduled.  Fortunately, the lack of a significant ramp at 4.5\,$\mu$m  allowed us to analyze the full data without an initial trim.  The eclipse {\bf is detected} at 4.5\,$\mu$m, but not at 3.6\,$\mu$m.

{\bf WASP-76b} required a dilution correction due to the presence of a close companion, entirely contained within {\it Spitzer's} point spread function. To calculate our dilution correction, we used the $\Delta{z}$ magnitude difference listed by \citet{wollert15}, and converted that to a difference in K-magnitude using Table~7 of \citet{covey07} under the assumption that both stars are on the main sequence.

{\bf WASP-103b} was analyzed by \citet{kreidberg18}, who derived quite a high value for the eclipse depth at 4.5\,$\mu$m ($5690\pm140$\,ppm).  We are skeptical that the eclipse depth can be that large, and we note that it was $2.9\sigma$ above their best-fit model.  Hence, we omitted the \citet{kreidberg18} measurement from the comparison in Figure~\ref{others_fig}.  However, our two values corrected for dilution ($5230\pm424$\,ppm and $5413\pm390$\,ppm) are in good agreement with their retrieved model (blue square on the right panel of their Figure~7).  Thus, we support their retrieved results for this planet. Our dilution correction is based on the K-magnitude difference from \citet{ngo16} and \citet{delrez18}. 

{\bf WASP-121b} was observed by \citet{evans17} and \citet{kovacs19} who quote 3.6\,$\mu$m {\it Spitzer} secondary eclipse values.  The preliminary depth and central phase values quoted by those authors were measured by one of us (D.D.), and are superceded by the final values in Tables~\ref{tdepth_table} and \ref{phase_table}.  (The differences between the preliminary and final values are minor.) \citet{kovacs19} derive an orbital eccentricity of $0.0207\pm0.0153$ based on timing and duration of the primary transit and secondary eclipse.  The prominent {\it Spitzer} eclipses (Figures~\ref{all36} and \ref{all45}) are very well fit using the orbital parameters derived for the transits by \citet{delrez16}.  Thus, we find no evidence for a difference in duration of the transit and eclipse.  Weighting the central phases of the two {\it Spitzer} bands by the inverse of their variance, and correcting for light travel time across the orbit, we find $e\,\cos{\omega} = -0.00088\pm0.00060$.

\end{document}